\documentclass[aps,prl,preprint,amsmath,amssymb,superscriptaddress]{revtex4-2}
\usepackage{graphicx}
\usepackage{dcolumn}
\usepackage{bm}

\usepackage{comment}
\usepackage{ulem}
\usepackage{soul}
\usepackage{xcolor}
\usepackage{amsmath} 
\usepackage{mathrsfs}
\usepackage{graphicx}
\usepackage{dcolumn}
\usepackage{bm}
\usepackage{anyfontsize}
\usepackage[utf8]{inputenc}
\usepackage[T1]{fontenc}
\usepackage{mathptmx}
\usepackage{etoolbox}
\usepackage{titletoc}
\usepackage{tocloft}
\usepackage[colorlinks=true, linkcolor=blue, citecolor=blue, urlcolor=blue, linktoc=all]{hyperref}

\begin{document}

\preprint{APS/123-QED}

\title[Sample title]{Ultrafast Critical Slowing of Spin Dynamics and Emergent Nonequilibrium Fano Interference in Fe$_3$GeTe$_2$ }
\date{\today}
             
 \author{Anupama Chauhan}
 \affiliation {Department of Physical Sciences, Indian Institute of Science Education and Research Kolkata, Mohanpur, West Bengal, 741246, India.}
\author{Sidhanta Sahu}
 \affiliation {Department of Physical Sciences, Indian Institute of Science Education and Research Kolkata, Mohanpur, West Bengal, 741246, India.}
 
 \author{Satyabrata Bera}
 \affiliation {School of Physical Sciences, Indian Association for the Cultivation of Science, Jadavpur, Kolkata, 700032, India.}
 
 \author{Tuhin Debnath}
 \affiliation {School of Physical Sciences, Indian Association for the Cultivation of Science, Jadavpur, Kolkata, 700032, India.}
 
\author{Mintu Mondal}
\affiliation {School of Physical Sciences, Indian Association for the Cultivation of Science, Jadavpur, Kolkata, 700032, India.}

\author{Anamitra Mukherjee}
\affiliation{School of Physical Sciences, National Institute of Science, Education and Research, HBNI, Jatni 752050, India}
\affiliation{Homi Bhabha National Institute, Training School Complex, Anushaktinagar, Mumbai 400094, India}

\author{Siddhartha Lal}
\affiliation {Department of Physical Sciences, Indian Institute of Science Education and Research Kolkata, Mohanpur, West Bengal, 741246, India.}

\author{N. Kamaraju}
\email{nkamaraju@iiserkol.ac.in} 
\affiliation {Department of Physical Sciences, Indian Institute of Science Education and Research Kolkata, Mohanpur, West Bengal, 741246, India.}
\altaffiliation {Department of Physical Sciences, Indian Institute of Science Education and Research Kolkata, Mohanpur, West Bengal, 741246, India.}

\begin{abstract}
Fe$_3$GeTe$_2$ is a prototypical metallic van der Waals ferromagnet with itinerant magnetism and a highly tunable Curie temperature, yet how electronic excitations couple to spin and lattice degrees of freedom across its magnetic transition remains largely unexplored. Here, we use two-color pump–probe reflectivity to investigate the coupled electronic, spin, and lattice dynamics. The time-resolved reflectivity exhibits a tri-exponential relaxation, in which the intermediate component shows an anomaly near the Curie temperature due to enhanced interlayer spin-lattice interactions, while the slowest component displays pronounced critical slowing down with an exponent of $\simeq$0.3, revealing non-universal relaxation dynamics associated with intralayer spin correlations. Furthermore, we observe an emergent nonequilibrium $A_{1g}$ phonon Fano asymmetry that is suppressed in the ferromagnetic phase but anomalously enhanced in the paramagnetic regime, driven by thermally activated anharmonic decay pathways that bridge the kinematic gap to a hot electronic continuum. The pronounced enhancement of the acoustic strain pulse amplitude near T$_c$ further evidences robust magnetoelastic coupling. Overall, our results reveal how magnetic order governs the interplay among critical spin dynamics, electronic continuum excitations, and lattice response in metallic van der Waals ferromagnets.

\end{abstract}

\maketitle

Low-dimensional itinerant ferromagnets provide an ideal platform to explore the interplay between electronic correlations, spin order, and lattice degrees of freedom, particularly in the vicinity of magnetic phase transitions. Among them, Fe$_3$GeTe$_2$ (FGT) has emerged as a model quasi-two-dimensional van der Waals (vdW) ferromagnet, exhibiting itinerant ferromagnetism\cite{Fei2018}, a highly tunable Curie temperature \cite{Liu2020,Deng2018}, large coercive fields \cite{Roemer2020}, and the stabilization of topological spin textures \cite{Wu2020,Birch2022}. In ambient conditions, FGT undergoes ferromagnetic ordering at T$_c\sim220$–230 K~\cite{Deiseroth2006,Calder2019}. 

Ultrafast optical spectroscopy provides direct access to the coupled electronic, spin, and lattice dynamics on their intrinsic timescales, and is therefore ideally suited to investigate demagnetization \cite{Beaurepaire1996,Lichtenberg2022}, magnetic critical dynamics \cite{Hohenberg1977,Lovinger2020_LVO,Zhou2022,Zhang2021,Sahu2025}, spin-lattice interactions \cite{Adhikari2025,Hu2023} and the dynamics of collective spin excitations in low-dimensional ferromagnets. By driving the system far from equilibrium, the femtosecond optical pulse excitation enables disentangling the effects of electron-phonon thermalization, spin-lattice relaxation, and the recovery of spin correlations, offering a powerful time-domain perspective on phase-transition dynamics \cite{Hohenberg1977}. While ultrafast critical slowing down has been extensively explored in antiferromagnetic \cite{Zhou2022,Zhang2021, Sahu2025} and charge-ordered systems \cite{Zong2019}, its manifestation in ferromagnetic metals, particularly in the time domain, remains far less investigated. The strongly itinerant and dissipative nature of ferromagnetic metals makes the identification of critical spin dynamics especially challenging, and direct experimental evidence is scarce.

In addition, the ultrafast optical excitation can generate non-equilibrium quantum interference between phonons with a discrete spectrum and electronic or spin-related continua, manifested as the Fano effect \cite{Fano1961,Klein1982}. In pump-probe\cite{Hase2005, *Hase2006, *Kumar2007, *Hase2003, *Lee2006} and Raman studies \cite{Misochko2000, *Tanwar2022}, such coupling produces an asymmetric phonon line shape characterized by the Fano parameter $q$, which is highly sensitive to changes in electronic coherence and spin correlations across phase transitions. We show here that in magnetic vdW systems such as FGT, where itinerant electrons, localized spins, and lattice vibrations are strongly intertwined, Fano interference provides a sensitive probe of phonon coupling to dissipative electronic continuum across the ferromagnetic (FM) to paramagnetic (PM) transition.

We conduct time-resolved studies using two-color pump-probe reflection spectroscopy to investigate the coupled dynamics of the electronic, spin, and phonon subsystems of FGT on ultrafast timescales. We identify three distinct relaxation channels associated with electron-phonon thermalization, interlayer spin-lattice relaxation, and intralayer spin-lattice dynamics in ascending timescales, and thereby observe pronounced critical slowing down of the slowest relaxation channel in the vicinity of T$_{c}$. Further, we uncover strong Fano interference between a coherent optical phonon and an electronic continuum above T$_{c}$, which weakens in the ferromagnetic phase. Microscopic modeling suggests that the temperature evolution of the Fano asymmetry originates from thermally-enhanced  overlap between the anharmonically broadened optical phonon and a strongly dissipative electronic continuum in the paramagnetic phase, while exchange splitting in the ferromagnetic phase suppresses the relevant low-energy electron–hole excitations, thereby weakening the continuum-mediated interference.
In addition, the amplitude of the strain pulse generated via thermoelastic and magnetic stress grows sharply near T$_c$, evidencing robust magnetoelastic coupling (see Fig.~1 for a summary of the salient observations).


\begin{figure}
    \centering
    \includegraphics[width=1\linewidth]{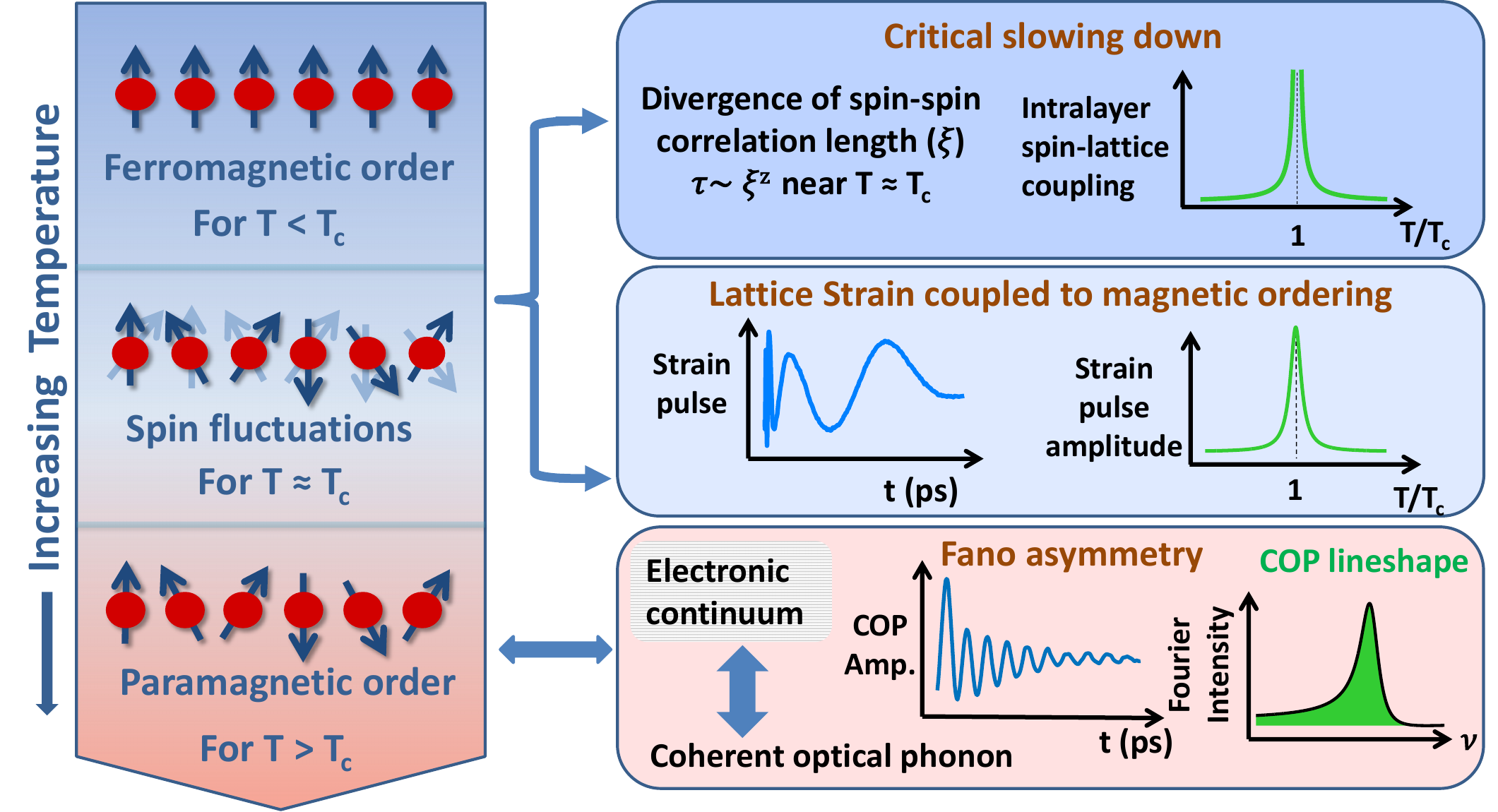}
    \caption{
Schematic illustrating the evolution of coupled electronic, spin, and lattice dynamics across the ferromagnetic phase transition in FGT. The left panel shows the temperature progression from the ferromagnetic phase (T < T$_c$) to the paramagnetic phase (T > T$_c$), with strong spin fluctuations near T $\approx$ T$_c$. The right panels summarize the dominant processes observed in this study: (top) critical slowing down of the intralayer spin-lattice relaxation associated with the divergence of the spin-spin correlation length near T$_c$; (middle) enhanced acoustic strain-pulse amplitude indicating strong magnetoelastic coupling; and (bottom) coupling between the electronic continuum and the coherent $A_{1g}$ optical phonon, producing a Fano-type asymmetric phonon lineshape above T$_c$.}
    \label{fig1}
\end{figure}

Our study investigates single-crystalline FGT using two-colour pump-probe reflection spectroscopy based on 250 kHz repetition rate femtosecond amplifier (details of sample preparation, characterization, and experimental setup are provided in the Supplemental Material (SM), Sec. I and II \cite{Supp}). Since the Fe-$3d$ orbitals have the highest density of states (DOS) \cite{Jiang2022, Zhuang2016}, the pump pulses (3.14 eV) primarily excite electrons from Fe-$3d$ states below the Fermi level into unoccupied Fe $3d$ states above the Fermi level, while the probe pulses (1.57 eV) are sensitive to Fe $3d$-$3d$ transitions across the Fermi level \cite{Jiang2022} (see SM, Sec. III \cite{Supp}). These excitations are consistent with Stoner-type interband transitions \cite{Zhuang2016}. Transient differential reflectivity ($\Delta$R/R) is measured over the temperature range 120–320 K with the pump and probe fluences fixed at 125 and 8~$\mu$J/cm$^{2}$, respectively.

\begin{figure}
    \centering
    \includegraphics[width=1\linewidth]{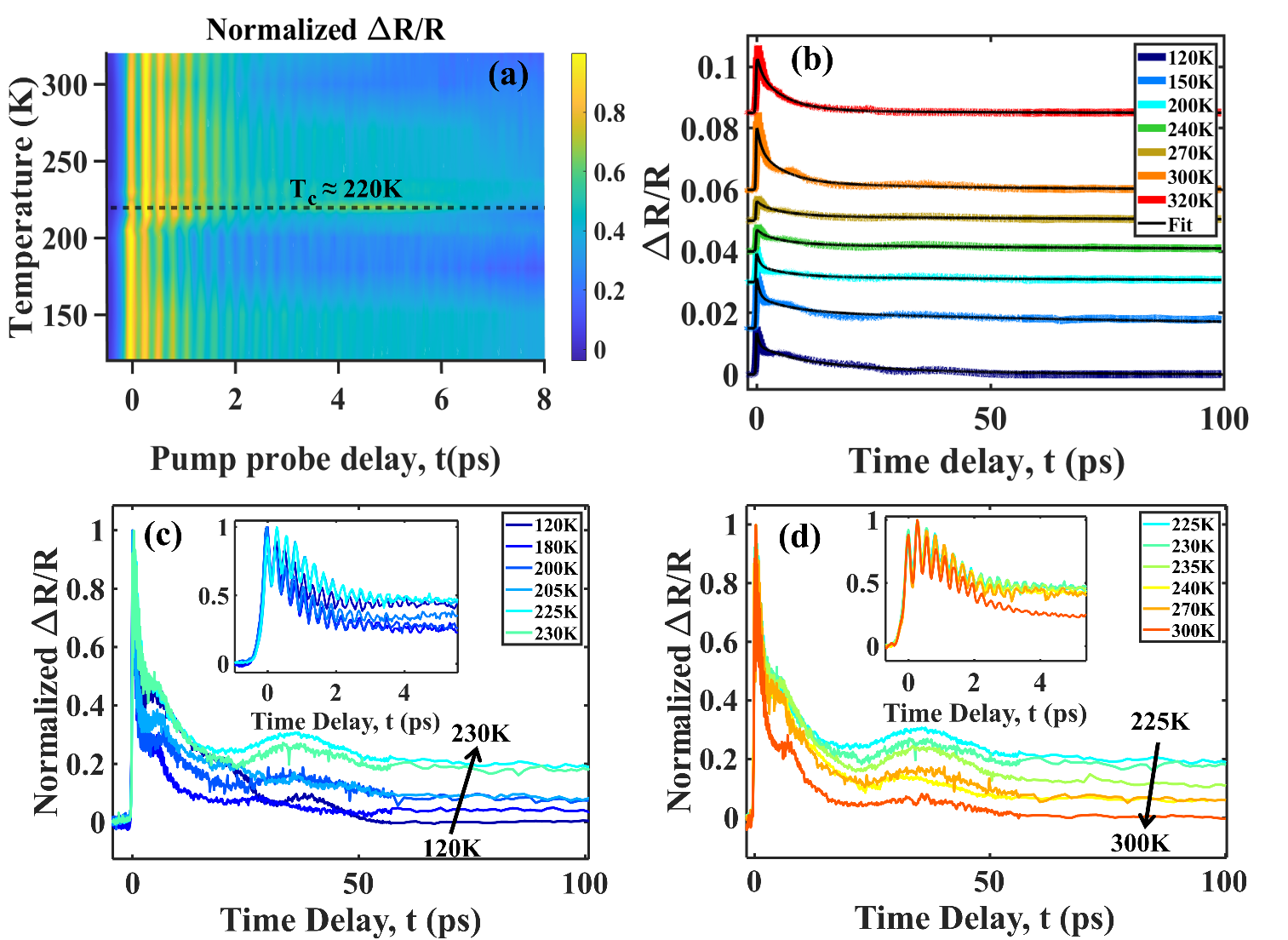}
    \caption{The time-resolved differential reflectivity measurements. (a) 2D false color plot of the normalized $\Delta$R/R vs temperature from 120-320 K. The black dashed line indicates the curie temperature (T$_{c}$). (b) The normalized $\Delta$R/R at few selected temperatures (different colors) and the fit with Eq. (\ref{eq1}) (black solid line). The data has been shifted vertically for clarity. (c) and (d) displays the variation in the exponential background below and above T$_{c}$ respectively. Inset in (c) and (d) show early-time dynamics upto 5 ps, highlighting the optical phonon oscillations. 
        }    \label{fig2}
\end{figure}

Figure~\ref{fig2}(a) presents a false-color map of the normalized $\Delta$R/R as a function of pump-probe delay (0-8 ps) and temperature (120--320 K). Near the ferromagnetic transition at 
$\sim$220 K, pronounced changes in the $\Delta$R/R are observed. The photoinduced $\Delta$R/R exhibits an initial positive response following excitation [Fig.~\ref{fig2}(b)], arising from the promotion of electrons to higher-energy states. These nonequilibrium carriers subsequently relax through energy exchange with the spin and lattice subsystems, giving rise to a multi-exponential decay of the background signal. Superimposed on this decay are two coherent contributions: a coherent optical phonon whose frequency matches the $A_{1g}$ Raman mode associated with out-of-plane vibrations of the Te atoms, and a propagating acoustic strain pulse.

The electronic background dynamics are well described by a tri-exponential function [black solid lines in Fig.~\ref{fig2}(b)] of the form
\begin{equation}
\frac{\Delta \text{R}}{\text{R}}= \frac{1}{2} \left[1+\text{erf} \left(\frac{t-t_{0}}{\tau_{r}}\right)\right]  \sum_{i=1}^{3} \text{A}_{i} e^{\text{-}t/\tau_{i}} ,
\label{eq1}
\end{equation}
where the error function accounts for the finite rise time 
$\tau_{r}\approx130$-200 fs. Here, A$_i$ and $\tau_i$ ($i$=1,2,3) denote the amplitudes and characteristic decay times of the three relaxation channels, respectively (see SM, Sec.~IV for details \cite{Supp}).

Figures~\ref{fig2}(c) and \ref{fig2}(d) show the temperature dependence of the normalized $\Delta$R/R, highlighting the evolution of the electronic background across T$_{c}$. The overall recovery time increases 
as the temperature approaches T$_{c}$ from below and decreases above T$_{c}$, consistent with critical slowing down near the magnetic phase transition.

\begin{figure}
    \centering
    \includegraphics[width=1.0\linewidth]{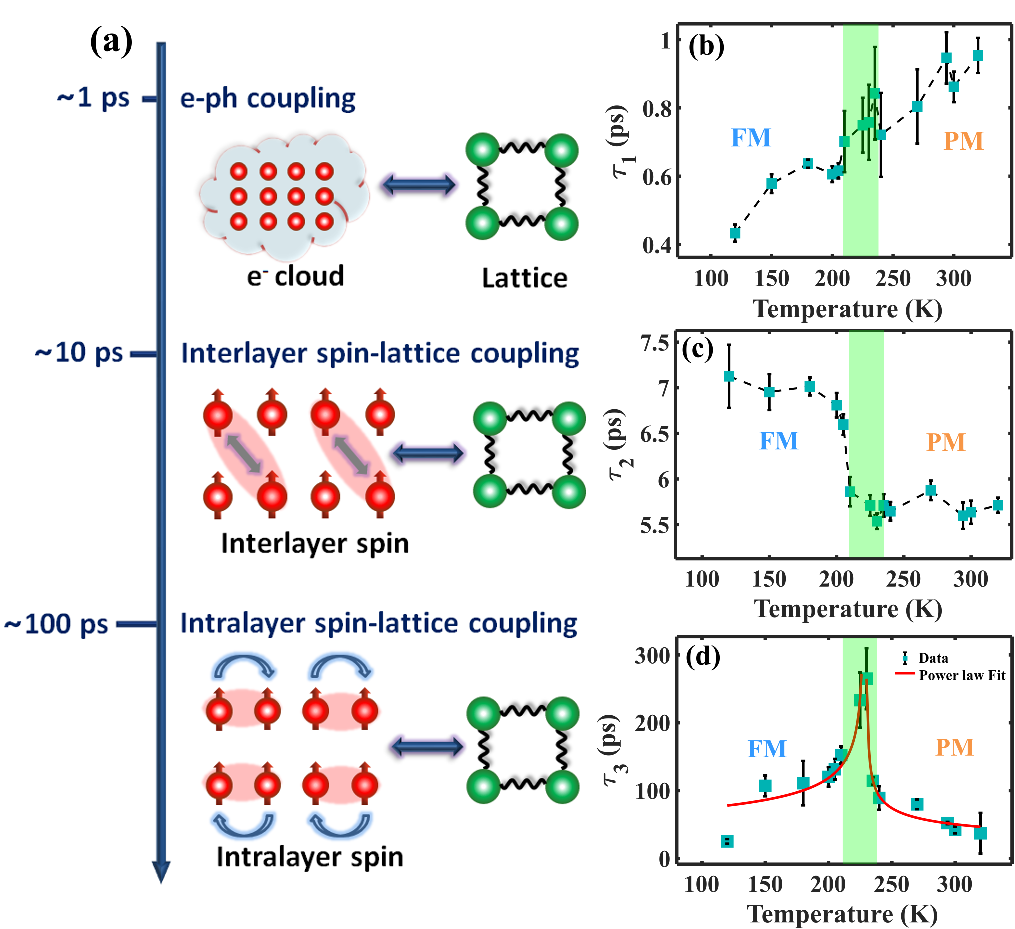}
    \caption{(a) Schematic depicting the hierarchy of relaxation processes, with electron- phonon coupling occurring on sub-ps timescales, followed by interlayer and intralayer spin- lattice relaxation on tens and hundreds of ps timescales. The relaxation timescales (b) $\tau_{1}$, (c) $\tau_{2}$ and (d) $\tau_{3}$ extracted from a triexponential fit to the $\Delta$R/R data (green squares) with error bars. The black dotted lines are guides to the eye. The light green shaded area highlights the region near the critical temperature, T$_{c}$. The red solid line in the (d) is the fit using the power-law (see text)}    \label{fig3}
\end{figure}

The temperature dependence of the decay time constants extracted from the fits to the electronic background is shown in Fig.~\ref{fig3}(a)-(c). Three distinct relaxation timescales are clearly resolved, $\tau_{1}\sim0.4$-1 ps, $\tau_{2}\sim5$-7 ps, and $\tau_{3}\sim40$-300 ps, indicating the presence of multiple relaxation channels governed by different microscopic processes. Upon optical excitation, the pump pulse drives Fe $3d$-$3d$ transitions \cite{Jiang2022,Zhuang2016}, generating a population of hot electrons. These carriers first undergo rapid electron-electron thermalization and subsequently relax by transferring energy to the 
lattice and spin subsystems \cite{Badrtdinov2023} [Fig.~\ref{fig3}(a)]. 

The fastest relaxation channel, $\tau_{1}$ [Fig.~\ref{fig3}(b)], lies in the sub-picosecond regime and increases monotonically from $\simeq$ 0.43 $\pm$ 0.02 ps at 120 K to $\simeq$ 0.95 $\pm$ 0.05 ps at 320 K. This behavior is characteristic of electron-phonon thermalization and is consistent with previous ultrafast studies on FGT \cite{Lichtenberg2022}.
After equilibration between the electron and phonon subsystems, further energy and angular momentum transfer proceeds via spin-lattice coupling. These processes are captured by the relaxation timescales $\tau_{2}$ and $\tau_{3}$, both of which exhibit pronounced changes across the magnetic transition temperature.
 
As the sample temperature is raised, the intermediate timescale, $\tau_{2}$, drops abruptly from $\simeq$ 7.1 $\pm$ 0.3 ps to $\simeq$ 5.7 $\pm$ 0.1  ps across T$_{c}$ beyond which it remains nearly temperature-independent [Fig. \ref{fig3}(c)]. 
Since long-range interlayer spin correlations are present only in the FM phase, the observed intermediate timescale is plausibly associated with interlayer spin-lattice dynamics. The reduction of $\tau_{2}$ upon warming through T$_c$ is consistent with the collapse of interlayer spin correlations and an increased contribution from electron-phonon scattering in the PM phase. This interpretation is consistent with optical and photoemission studies of Fe$_3$GeTe$_2$, which reveal strong electronic correlations and a temperature-driven incoherent-to-coherent crossover characterized by spectral weight transfer and the emergence of quasiparticle coherence upon cooling~\cite{Corasaniti2020, Sharma2024}.

Further insight is provided by angle-resolved photoemission spectroscopy measurements by Xu \textit{et al.}~\cite{Xu2020}, which reveal a strong temperature evolution of the electronic structure in Fe$_3$GeTe$_2$ and point to a correlation-driven incoherent-to-coherent electronic crossover upon cooling. Within this framework, the emergence of more coherent charge carriers for T < T$_c$ is consistent with reduced magnetic-fluctuation-mediated scattering, leading to longer $\tau_{2}$. Conversely, for T > T$_c$, enhanced magnetic fluctuations together with incoherent electronic behavior are expected to increase available scattering channels, resulting in faster relaxation.
The characteristic timescale of order $\sim 5-7$~ps for $\tau_{2}$ is comparable to the demagnetization time reported in Fe$_3$GeTe$_2$~\cite{Lichtenberg2022} and to spin-lattice relaxation times observed in rare-earth ferromagnets such as Gd and Tb alloys~\cite{Eschenlohr2014, Andres2021}. 

The slowest relaxation channel, $\tau_{3}$, spans approximately $40$-$300$~ps and exhibits a pronounced increase upon approaching T$_c$ from both sides [Fig.~\ref{fig3}(d)], consistent with critical slowing down near a continuous magnetic phase transition (see SM, Sec. VI \cite{Supp}, for the robustness of $\tau_3$ fitting). This component is associated with the recovery of short-range intralayer spin correlations that are strongly coupled to the lattice. Such correlations arise from the combined effects of intra- and interlayer exchange interactions within the Fe-Te planes (see SM, Sec.~III \cite{Supp}). A phenomenological description of the observed critical behavior of $\tau_{3}$ is presented in the End Matter (see SM, Sec. V \cite{Supp} for detailed derivation).

\textcolor{black}{The temperature dependence of $\tau_{3}$ is well described by a power-law form \cite{Zhou2022}, $\tau_3=\tau_{0}(1-\text{T/T}_{c})^{-m}$, where $\tau_{0}$ is the critical amplitude and $m$ is the dynamical critical exponent}. Below T$_{c}$, the fit yields $\tau_0^-= 62\pm10$ ps, T$_{c}=228\pm3$ K, and $m=0.31\pm0.04$, while above T$_{c}$ we obtain $\tau_0^+= 35\pm5$ ps, T$_{c}=230\pm3$ K, and $m=0.30\pm0.05$. For a conventional second-order phase transition, the critical amplitude above T$_c$ is expected to be approximately twice that below T$_c$ (see End Matter). In contrast, our analysis reveals the opposite trend, indicating that the relaxation dynamics associated with $\tau_{3}$ deviate from simple universal scaling behavior \cite{Hohenberg1977}.

Consistently, the extracted dynamical critical exponent $m \simeq 0.30$ is significantly smaller than the universal values ($m \sim 1.2$-$1.4$) predicted for standard magnetic models~\cite{Hohenberg1977, Peczak1993, Wansleben1991}. The combination of non-universal critical amplitudes and an unusually small exponent suggests that the dynamics underlying $\tau_{3}$ are influenced by material-specific interactions beyond long-wavelength critical fluctuations alone. In particular, local spin-lattice coupling, magnetic anisotropy, and disorder are expected to play an important role in renormalizing the effective critical dynamics. Comparable small values of the dynamical critical exponent ($m \sim 0.2$--$0.4$) have been reported for the slowest relaxation channel across the phase transition in LaVO$_3$~\cite{Lovinger2020_LVO}.

We now turn to the coherent optical phonon response in the $\Delta$R/R signal. Fe$_3$GeTe$_2$ crystallizes in the hexagonal P6$_3$/mmc structure, consisting of Fe$_3$Ge slabs separated by van der Waals bonded Te layers, and supports a fully symmetric $A_{1g}$ optical phonon mode. The generation of coherent optical phonons can be understood within the framework of displacive excitation of coherent phonons (DECP)~\cite{Zeiger1992, Merlin1997}, which is expected to dominate in strongly absorbing materials following ultrafast photo-excitation. Recent nonequilibrium TDDFT studies further highlighted the important role of coherent $A_{1g}$ phonons in ultrafast spin and electronic dynamics in FGT~\cite{Wu2026}.

\begin{figure}
    \centering
    \includegraphics[width=1.0\linewidth]{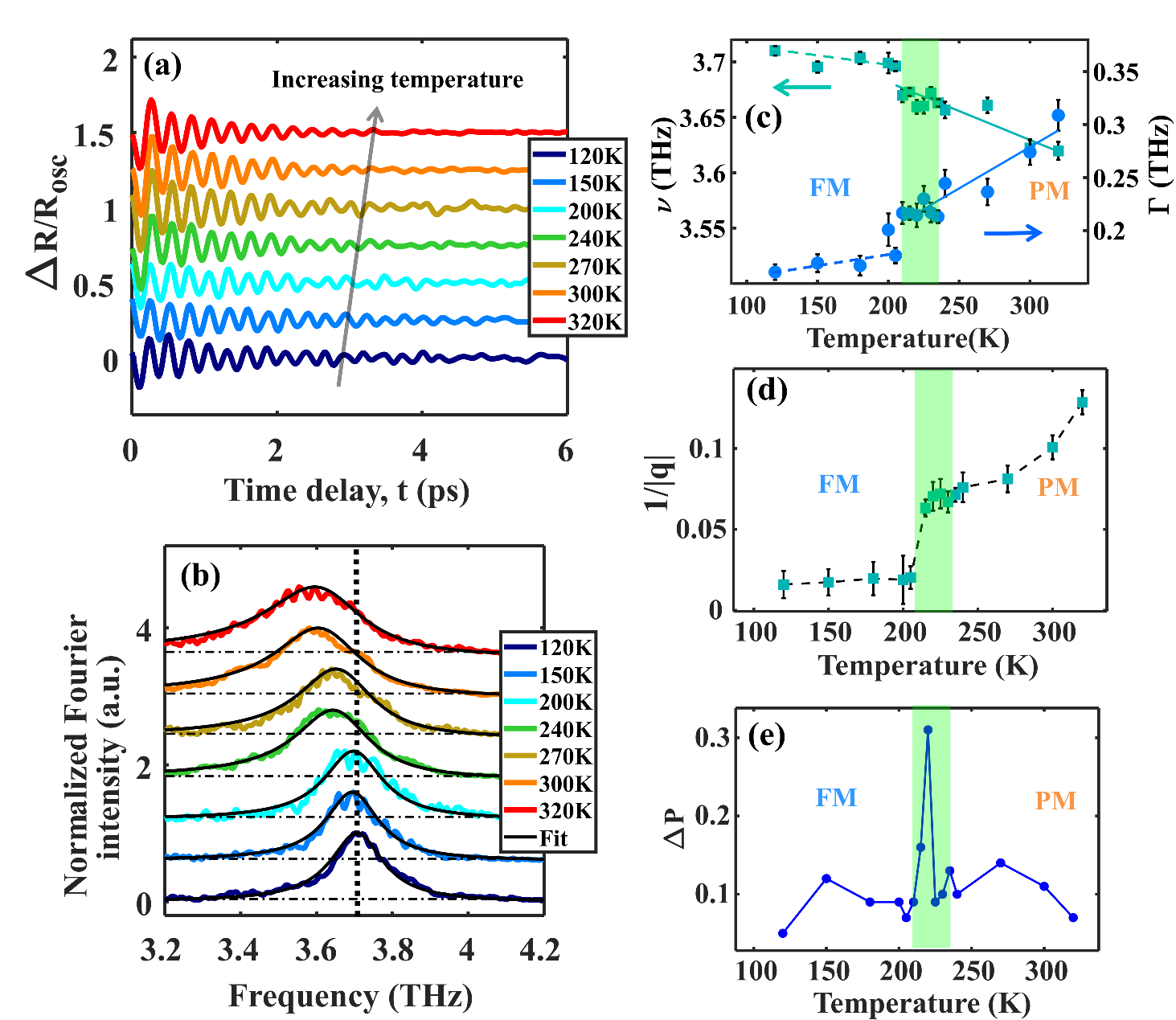}
    \caption{ (a) The oscillatory component (A$_{1g}$ optical phonon) extracted from the transient normalized $\Delta$R/R signal and the corresponding (b) Normalized Fourier intensity spectra at few selected temperatures; the solid black curve represents a fit to the Fano line shape (Eq. \ref{eq2}). The arrow denotes increasing temperature and the black dotted lines serve as guide to the eye. Temperature dependence of the phonon frequency ($\nu$, left axis) and linewidth ($\Gamma$ , right axis) obtained from the Fano fits. The solid green and blue curves show the fit to the cubic anharmonic decay model in the high-temperature regime (210-320 K), while the dashed lines are guides to the eye. (d) Temperature evolution of the Fano asymmetry parameter ($1/|q|$). (e) The strain component is extracted from the transient $\Delta$R/R signal and it's amplitude (dip-to-peak difference, ($\Delta$P)) is plotted as a function of temperature (see SM, Sec. IX, for details \cite{Supp}). The light green shaded region marks the vicinity of the Curie temperature, T$_{c}$. 
}    \label{fig4}
\end{figure}

The oscillatory phonon component is isolated by subtracting the triexponential electronic background and removing the strain-pulse contribution (using Butterworth filter in MATLAB) at each temperature. The resulting $\Delta$R/R$_{\mathrm{osc}}$ and corresponding Fourier spectra are shown in Figs.~\ref{fig4}(a) and \ref{fig4}(b). With increasing temperature, the $A_{1g}$ phonon exhibits a systematic redshift and linewidth broadening. Notably, above $\sim215$~K, the phonon spectra develop a pronounced asymmetric Fano lineshape, a feature absent in Raman measurement~\cite{Du2017}. The spectra are well fitted by a Breit-Wigner-Fano function~\cite{Fano1961, Hase2006},
\begin{equation}
\text{I} = \text{I}_{0}\frac{(1+\epsilon/q)^{2}}{1+\epsilon^{2}}, \label{eq2}
\end{equation}
where $\epsilon = (\nu-\nu_{0})/(\Gamma/2)$, $\nu_{0}$ is the bare phonon frequency, $\Gamma$ is the linewidth of the phonon, and $q$ is the asymmetry parameter. 

Above T$_c$, in the PM phase, both the phonon frequency $\nu$ and linewidth $\Gamma$ follow a cubic anharmonic decay (blue and green solid  lines) [Fig.~\ref{fig4}(c)]. The details of the fitting using this model are provided in the SM, Sec.~VII \cite{Supp}. The fit yields $\nu_{0} = 3.78 \pm 0.02$~THz, $A = -0.024 \pm 0.004$~THz, $\Gamma_{0} = 0.04 \pm 0.03$~THz, and $C = 0.036 \pm 0.005$~THz. In contrast, below T$_c$, both $\nu$ and $\Gamma$ in the ferromagnetic phase (blue and green dashed lines) deviate from the cubic anharmonic behavior, indicating an additional contribution beyond pure anharmonicity \cite{Tian2016,Du2017}.

We next examine the Fano asymmetry parameter $1/|q|$, which is negative, indicating asymmetry toward lower frequencies. In Fe$_3$GeTe$_2$, $1/|q|$ remains nearly temperature independent in the ferromagnetic phase but increases strongly with temperature in the paramagnetic phase [Fig.~\ref{fig4} (d)]. Similar asymmetric phonon responses have been reported for the coherent $A_{1g}$ mode in cuprate superconductors \cite{Misochko2000} and the $E_{2g}$ mode in Zn \cite{Hase2006}. Since long-wavelength magnons exist only in the ferromagnetic phase~\cite{Trainer2022, Bansal2023, Calder2019}, the enhanced Fano asymmetry above T$_c$ excludes magnon--phonon coupling as the dominant mechanism and instead points to coupling between the discrete $A_{1g}$ phonon and an electronic continuum.

In general, for metals, the Fano lineshape of the Raman phonons has been theoretically ascribed to contributions from both interband and intraband electronic Raman scattering \cite{Klein1982}. In FGT, the interband Fe $3d$-$3d$ transition excited by the 3.14 eV pump along the $\Gamma'$–K direction likely contributes to the electronic continuum \cite{Jiang2022}. 

The enhanced Fano asymmetry in the paramagnetic phase is understood within a microscopic electron-phonon framework (see End matter and SM, Sec. X \cite{Supp}). Here, the anharmonic broadening and softening of the coherent $A_{1g}$ phonon facilitate its overlap with a strongly dissipative electronic continuum generated by pump-induced hot carriers \cite{Rath2025,Cheng2025}. Crucially, thermal acoustic phonons act as an efficient momentum bridge, bypassing the kinematic constraints of the zone-center optical probe. As a result, the magnitude of the Fano asymmetry ($1/|q|$) increases monotonically with lattice temperature in the PM phase, driven by the Bose distribution-enhanced acoustic phonon population maximizing the phase-space overlap between the discrete optical mode and the smeared electron-hole continuum. Conversely, in the ferromagnetic phase, the onset of a large exchange splitting strongly quenches this scattering channel. The relevant low-energy electron-hole transitions across the split bands necessitate a spin-flip---a requirement the non-magnetic $A_{1g}$ optical phonon cannot satisfy due to angular momentum conservation, thereby restoring a symmetric Lorentzian phonon response. This mechanism is further corroborated by fluence-dependent measurements, which reveal a nearly linear increase of $1/|q|$ with pump fluence (see SM, Sec.~VIII \cite{Supp}), consistent with the direct enhancement of the hot electronic continuum at higher excitation densities.

The strain pulse generated by the pump excitation displays a distinct temperature dependence [Fig.~\ref{fig4} (e)]. While its amplitude remains nearly unchanged over a broad temperature range, it exhibits a marked enhancement in the vicinity of the ferromagnetic transition. This behavior signifies strong coupling between the transient lattice strain and magnetic order, providing clear evidence of significant magnetoelastic interactions that dynamically link the spin and lattice subsystems \cite{Lovinger2020}. On the other hand, these strain pulses scale linearly with pump fluence indicating the dominance of thermoelastic stress \cite{Thomsen1986} in the PM region. See SM, Sec. IX \cite{Supp} for the detailed discussion on the behaviors of these strain pulses with temperature and the pump fluence.

In conclusion, this study reveals the critical interplay between electronic, spin, and lattice degrees of freedom across the magnetic phase transition of the van der Waals ferromagnet FGT. The tri-exponential relaxation of the transient reflectivity resolves distinct interlayer and intralayer spin-related channels, with the latter exhibiting pronounced critical slowing down near T$_c$. The extracted power-law exponent for $\tau_3$, $m \simeq 0.3$, indicates non-universal critical dynamics associated with intralayer spin correlations rather than conventional magnetic universality. In parallel, the coherent $A_{1g}$ optical phonon displays strong Fano interference in the paramagnetic phase, arising from coupling to an electronic continuum, as further supported by the monotonic increase of the Fano asymmetry with pump fluence. Upon entering the ferromagnetic phase, the phonon lineshape evolves toward a nearly Lorentzian profile. Microscopic modeling suggests that this behavior originates from enhanced overlap between anharmonically broadened optical phonons and a strongly dissipative electronic continuum in the paramagnetic phase, whereas exchange splitting suppresses the relevant low-energy electron–hole excitations in the ferromagnetic state. The enhanced amplitude of the acoustic strain pulse near T$_c$ provides additional evidence for strong magnetoelastic coupling in the vicinity of the magnetic transition. Our results reveal how magnetic order controls the interplay between critical spin dynamics, dissipative electronic continua, and coherent lattice excitations in itinerant van der Waals ferromagnets.

 The authors thank the Ministry of Education (MoE), Government of India, for financial support and IISER Kolkata for providing the infrastructure to carry out this research.The authors acknowledge Poulami Ghosh and Pedisetti Venkatesh for their assistance during the experiments. The authors thank Prof. Venkataramanan Mahalingam for generously providing cryostat during the measurements. A.C. and S.S. acknowledge CSIR and UGC, respectively, for research fellowships. A.M., S.L. and N.K. thank ANRF, Govt. of India for funding through Research Grant ANRF/ARG/2025/004414/PS.



\newpage
\setcounter{equation}{0}
\renewcommand{\theequation}{\roman{equation}}
\onecolumngrid

\section{Appendix}

\textbf{\textit{Phenomenological Ginzberg-Landau Formalism:}}
For a second-order phase transition, the Landau free energy can be written as \cite{Khomskii2010}:

\begin{equation}
    f(T)= f_0(T)+ \alpha(T-T_c)p^2+\frac{1}{2}\beta p^4, \label{eq_i}
\end{equation}

where $p$ is the order parameter, and $\alpha$ and $\beta$ are Landau coefficients with $\beta>0$. The term $f_0(T)$ represents the temperature-dependent background of the high-temperature phase in the vicinity of the transition.

The order-parameter dynamics are then described within the relaxational (Model A) framework \cite{Hohenberg1977},

\begin{equation}
    \frac{1}{\mathscr{D}}\frac{\partial p}{\partial t}= -\frac{\partial f}{\partial p} \label{eq_ii}
\end{equation}

where $\mathscr{D}$ denotes the kinetic coefficient governing the relaxation of $p$. In the linear response regime, substituting the relaxation dynamics at equilibrium ($p=p_0$) back to Eq. \ref{eq_ii} results in the following form, 
\begin{equation}
    \delta p(t)= \delta p_0 ~e^{-t/\tau}, 
~~~where~~~ 
        \frac{1}{\tau}=-\mathscr{D}\frac{\partial^2 f}{\partial p^2}\biggr|_{p_0}\label{eq_viii}
\end{equation}

Therefore, $\tau$ can be estimated from the second-order derivative of free energy using the equilibrium value of $p_0$ below and above T$_c$ as follows:

\begin{equation}
\tau=\begin{cases} \frac{1}{2\alpha\mathscr{D}(T-T_c)} & \text{for } T>T_c \\
\frac{1}{4\alpha\mathscr{D}(T_c-T)} & \text{for } T<T_c\end{cases}
\end{equation}
 
Here, $(4\alpha\mathscr{D})^{-1}=\tau_{0}^-$ and $(2\alpha\mathscr{D})^{-1}=\tau_{0}^+$. The detailed derivation is present is the SM, Sec. V \cite{Supp}.
Thus, $\tau$ exhibits critical slowing down in the vicinity of T$_c$, which is experimentally manifested in the slowest relaxation component, $\tau_3$. The corresponding critical amplitudes above and below T$_c$, denoted as $\tau_0^{+}$ and $\tau_0^{-}$, respectively, define the amplitude ratio $R_{\tau} = \tau_0^{+}/\tau_0^{-}$. While conventional scaling predicts $R_{\tau} \simeq 2$, we instead obtain $R_{\tau} \simeq 0.6$ (see main text). This anomalous amplitude ratio indicates that the relaxation dynamics associated with $\tau_3$ deviates from simple universal scaling behavior.

\textbf{\textit{Thermal phonons-mediated Fano asymmetry:}} For understanding the Fano resonance in Fe$_3$GeTe$_2$, we employ a two-temperature model. Following ultrafast excitation, a strict temporal hierarchy justifies our quasi-equilibrium two-temperature model. Electrons rapidly thermalize within $\tau_{e-e} \sim 10\text{--}250~\text{fs}$  \cite{Beaurepaire1996, Beaurepaire1998} to an effective temperature (T$_e$), while lattice heating (T$_p$) via electron-phonon transfer takes few picoseconds ($\tau_{e-p}$)  \cite{Lichtenberg2022,Gou2025}. Because the $A_{1g}$ coherent phonon period ($\nu \approx 3.7~\text{THz}$, $\tilde{t}_{\text{phonon}} \approx 270~\text{fs}$) sits between these extremes ($\tau_{e-e} \ll \tilde{t}_{\text{phonon}} \ll \tau_{e-p}$), the hot electron bath acts as a stationary, thermalized continuum during coherent electron-phonon scattering. 

\begin{figure}[t]
    \centering
    \includegraphics[width=1.0\linewidth]{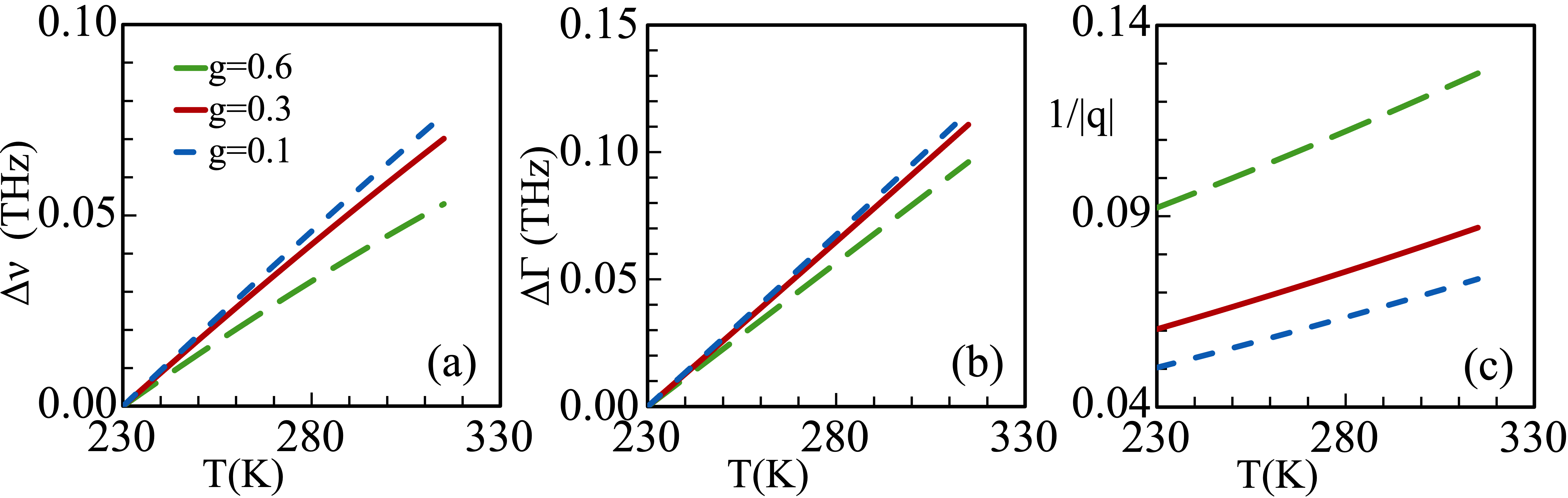}
    \caption{\textbf{Thermal phonons boosted Fano asymmetry:} Calculated softening, $\Delta \nu (T_p)=\nu(T_c)-\nu(T_p)$, broadening $\Delta \Gamma (T_p)=\Gamma(T_p)-\Gamma (T_c)$ and Fano asymmetry parameter $1/|q|$ as a function of the phonon temperature (T$_p$) in the paramagnetic regime (T > T$_c$).}
    \label{sfig1}
\end{figure}

We describe the electron-hole continuum at T$_e$ via the retarded electronic polarization bubble, $\Pi^R_0(\mathbf{q}, \nu, T_e)$. The phonon, at temperature T$_p$, is represented by the dressed retarded propagator $D^R(\mathbf{q}, \nu, T_p)$, which includes self-energy contributions from the anharmonic decay of optical phonons into pairs of acoustic phonons with opposite momenta ($\nu_0/2$). Following the Klemens model \cite{Cardona1983,Klemens1966}, this acoustic phonon loop self-energy grows with increasing T$_p$; its real part softens the optical phonon frequency, while its imaginary part induces broadening. To relax via the phonon bath, the excited electronic system requires both a finite electron-phonon matrix element coupling $\Pi^R_0$ to $D^R$, and an on-shell overlap in their respective densities of states. In the paramagnetic phase, the pump-induced hot electron temperature (T$_e$) heavily smears the Fermi surface, establishing a finite electronic continuum ($\mathrm{Im}[\Pi^R_0]$) that partially lifts zero-temperature kinematic constraints. Crucially, as T$_p$ monotonically increases, the Klemens-induced frequency softening actively drives the phonon deeper into this generated electronic continuum. This dynamic enhancement of the phase-space overlap directly drives the temperature-dependent Fano asymmetry.

In our modeling, detailed in the SM, Sec. X \cite{Supp}, we employ a general electron-phonon model comprising electrons with a parabolic dispersion, a zone-center optical phonon with a bare frequency of $\nu_0$, and acoustic phonons at $\nu_0/2$. We allow both types of phonons to interact with the electrons. By formally integrating out the acoustic phonons and setting the probe momentum ($\mathbf{q}_{\mathrm{pr}}$) close to the zone center, the microscopic optical response (phonon lineshape) is analytically derived as:

\begin{equation}
\begin{aligned}
I(\mathbf{q}_{\mathrm{pr}},\nu,T_p,T_e)
\propto
\operatorname{Im}\Bigg[
&\mathcal{M}_e^2 \Pi_0^R
+ \frac{
D^R
\left(
\mathcal{M}_p
+ g \mathcal{M}_e \Pi_0^R
\right)^2
}{
1 - g^2 D^R \Pi_0^R
}
\Bigg],
\end{aligned}
\label{eq:fano}
\end{equation}

Here, $D^R \equiv D^R(\mathbf{q}_{\mathrm{pr}},\nu,T_p)$ and 
$\Pi_0^R \equiv \Pi_0^R(\mathbf{q}_{\mathrm{pr}},\nu,T_e)$.
while $\mathcal{M}_{e}$ and $\mathcal{M}_{p}$ are the respective bare electronic and phonon dipole matrix elements, and $g$ represents the electron-phonon coupling, between the optical phonons and the electronic continua. The Klemens model parameters for softening and broadening are chosen to mimic the experiment. For parameters relevant to FGT (see, SM, Sec. X \cite{Supp}), and dimensionless dimensionless coupling strengths  $g\in[0.1, 0.6]$, in Fig. 5, we extract and plot, the softening, broadening and $1/|q|$ (the Fano asymmetry parameter) as a function of the sample temperature, identified with T$_p$. We notice that the Fano asymmetry increases with T$_p$ monotonically, while increase in electron-phonon  coupling raises the overall magnitude.

\newpage



\clearpage

\begin{center}
    \Large \textbf{Supplemental Material}
\end{center}

\renewcommand{\thesection}{\Roman{section}}
\renewcommand{\thepage}{S\arabic{page}}

\setcounter{figure}{0}
\renewcommand{\figurename}{Figure}
\renewcommand{\thefigure}{S\arabic{figure}}

\setcounter{table}{0}
\renewcommand{\tablename}{Table}
\renewcommand{\thetable}{TS\arabic{table}}




\section*{Table of Contents}
\startcontents
\setcounter{page}{1}
\printcontents{0}{0}{}




\newpage
\section{Sample Preparation and Characterization}
 
Single crystals of Fe$_{3}$GeTe$_{2}$ (FGT) were synthesized using the chemical vapor transport (CVT) method with I$_2$ as the transport agent. High-purity elemental powders of Fe (99.99 $\%$ pure), Ge (99.99 $\%$ pure), and Te (99.99 $\%$ pure) were mixed in a stoichiometric molar ratio of 3:1:2, thoroughly ground, and combined with 5 mg cm$^{-3}$ of I$_2$. The mixture was sealed under a vacuum of 2 $\times$ 10$^{-4}$ mbar in a quartz tube and placed in a two-zone horizontal furnace for seven days, with the hot and cold ends maintained at 750$^\circ$C and 700$^\circ$C, respectively. Shiny plate-like single crystals of approximate dimensions 2 $\times$ 1.5 $\times$ 0.1 mm$^{3}$ were harvested from the cold end. 

\begin{figure}
    \centering
    \includegraphics[width=1.0\linewidth]{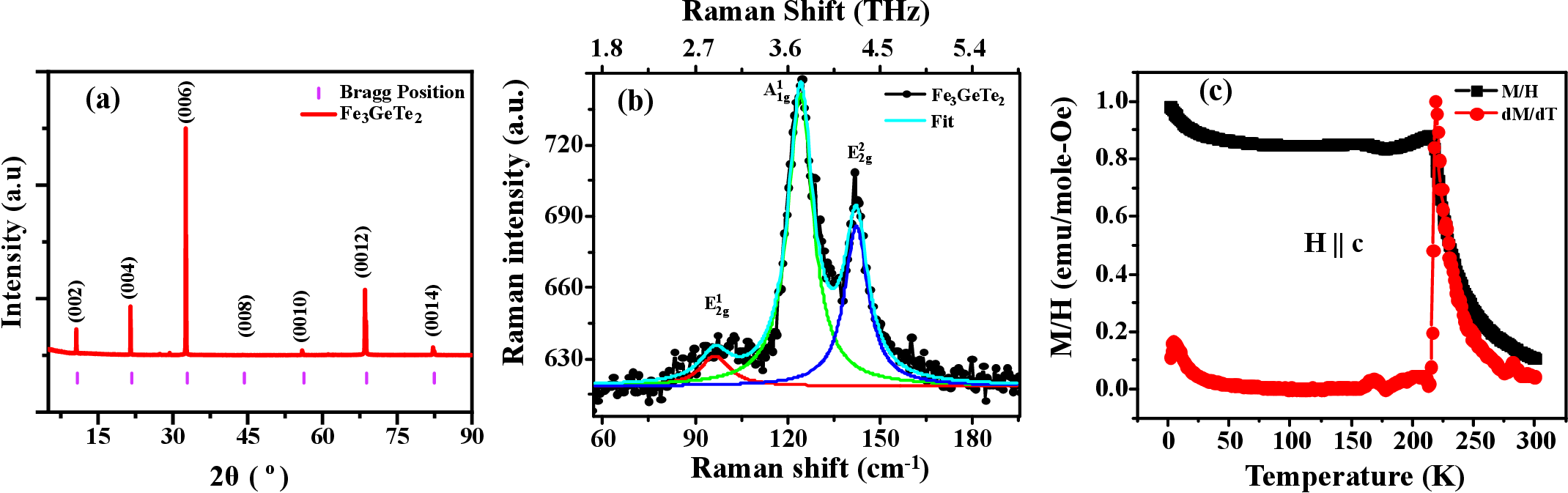}
    \caption{(a) X-ray diffraction spectrum of a single crystal of FGT, showing peaks corresponding to the ($00l$) planes, here $l=$ 2. (b) The Raman spectrum of bulk FGT (black line) is presented together with its multi-peak fit (cyan line). The corresponding multi-Lorentzian fit shown in the cyan line. The spectrum is well reproduced using three Lorentzian components, shown individually by the red, green, and violet curves, highlighting the distinct vibrational modes characteristic of the crystal structure. (c) The temperature-dependent magnetization measured under a magnetic field of 0.5 kOe applied along the c-axis in the field-cooled configuration. The magnetization (black squares) shows the expected ferromagnetic transition, while the derivative dM/dT (red circles) exhibits a pronounced peak at $\sim$ 218 K, identifying the Curie temperature of the sample.}
      \label{fig_S1_supplementary}
\end{figure}

The single crystal growth was characterized by X-ray diffraction (XRD), Raman spectroscopy and magnetization versus temperature (M-T) measurements. The crystal structure was confirmed by X-ray diffraction (XRD) using a Rigaku diffractometer with Cu-$K_{\alpha}$ radiation ($\lambda$ = 1.54056 Å), whose diffraction peaks corresponded to the expected Bragg reflections and confirmed c-axis orientation [see Fig. \ref{fig_S1_supplementary}(a)].

Raman spectroscopy was performed using a Horiba Jobin Yvon HR800 system with 488 nm laser excitation. The Raman spectrum, along with a multiple Lorentzian fit, is presented in Fig. \ref{fig_S1_supplementary}(b). We observe three distinct phonon modes i.e. E$_{2g}^{1}$, A$_{1g}^{1}$ and E$_{2g}^{2}$ (see Table \ref{Table1}), in good agreement with previous reports \cite{Du2017, Weerahennedige2024, Hu2023_2, Bera2023}. We note that the A$_{1g}$ phonon mode (124.2 cm$^{-1}$/3.73 THz) follows a Lorentzian lineshape in contrast to our observation in time-resolved measurements.

Magnetization (M-T) measurements were done using the MPMS-SQUID system in the 2–300 K temperature range, with care taken to align the plate-like crystals to minimize orientation errors. The magnetization was measured as a function of temperature under a magnetic field of 0.5 kOe applied along the c-axis. Fig. \ref{fig_S1_supplementary}(c) displays the magnetization and its temperature derivative, the latter exhibiting a pronounced peak at $\sim$218 K that defines the Curie temperature of the sample.

\begin{table}
\caption{\label{tab:table2}Fitted-Lorentzian parameters of the observed Raman modes}
\begin{tabular}{|c|c c c|}
\hline
 & E$_{2g}^{1}$& A$_{1g}^{1}$  & E$_{2g}^{2}$   \\
\hline

Raman shift (cm$^{-1}$) & 96.4 $\pm$ 0.9 & 124.2 $\pm$ 0.1 & 142.3 $\pm$ 0.2\\
Linewidth (cm$^{-1}$) & 11.6 $\pm$ 3.1   &	9.8 $\pm$ 0.3 & 9.4 $\pm$ 0.5\\
\hline

\end{tabular}\label{Table1}
\end{table}

\section{Experimental Setup}

\begin{figure}
    \centering
    \includegraphics[width=1.0\linewidth]{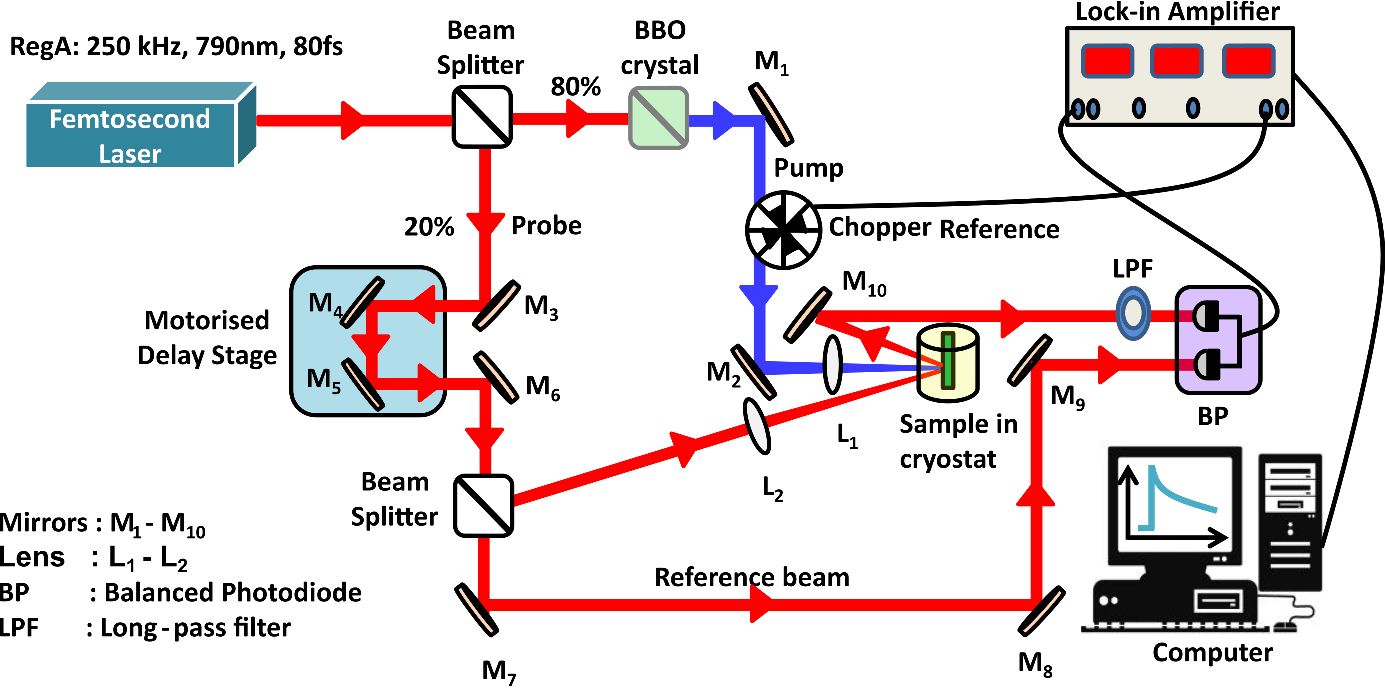}
    \caption{Schematic diagram of two-colour pump probe spectroscopy setup in reflection geometry.}
      \label{fig_S2_supplementary}
\end{figure}

Time-resolved reflectivity measurements were carried out using a home-built two-colour pump-probe setup, with a Regenerative femtosecond Amplifier (RegA 9050, Coherent Inc.) operating at a central wavelength of 790 nm and a repetition rate of 250 kHz, delivering an average power of 1.6 W. The laser produced 80 fs pulses with an energy of 6.4 $\mu$J. The output beam from the RegA was split into two parts by an 80:20 beam splitter. The higher-intensity portion was frequency-doubled to 395 nm using a $\beta$-barium borate crystal to serve as the pump, while the remaining fundamental beam was used as the probe. A variable delay stage is employed to introduce the time delay between the pump and probe pulses at the sample location. The sample is mounted in a liquid nitrogen cooled cryostat (Janis Inc. USA) and studied within a temperature range of 120 K to 320 K. The transient change in the probe beam's reflectivity induced by the pump pulse is measured using a balanced photo-diode. A lock-in amplifier is used to achieve phase-sensitive detection with an optical chopper modulating the pump beam at 711 Hz. To prevent scattered pump light from reaching the detector (photo-diode), a long-pass filter is used to block the pump light at 395 nm while allowing the probe at 790 nm to pass through. The pump and probe FWHM diameters are roughly $\sim$ 52 $\mu$m and $\sim$ 48 $\mu$m, respectively. The pump fluence is varied from 62 $\mu$J/cm$^{2}$ to 484 $\mu$J/cm$^{2}$ while the probe fluence is kept constant at 8 $\mu$J/cm$^{2}$ for the fluence dependent measurements. The instrument's response function, measured by the cross-correlation between the pump and probe pulses at the sample position, is $\sim$ 112 fs in our experiments. The experimental setup is shown in Fig. \ref{fig_S2_supplementary}.

\section{Exchange Interactions in FGT}

 \begin{figure}
    \centering
    \includegraphics[width=0.9\linewidth]{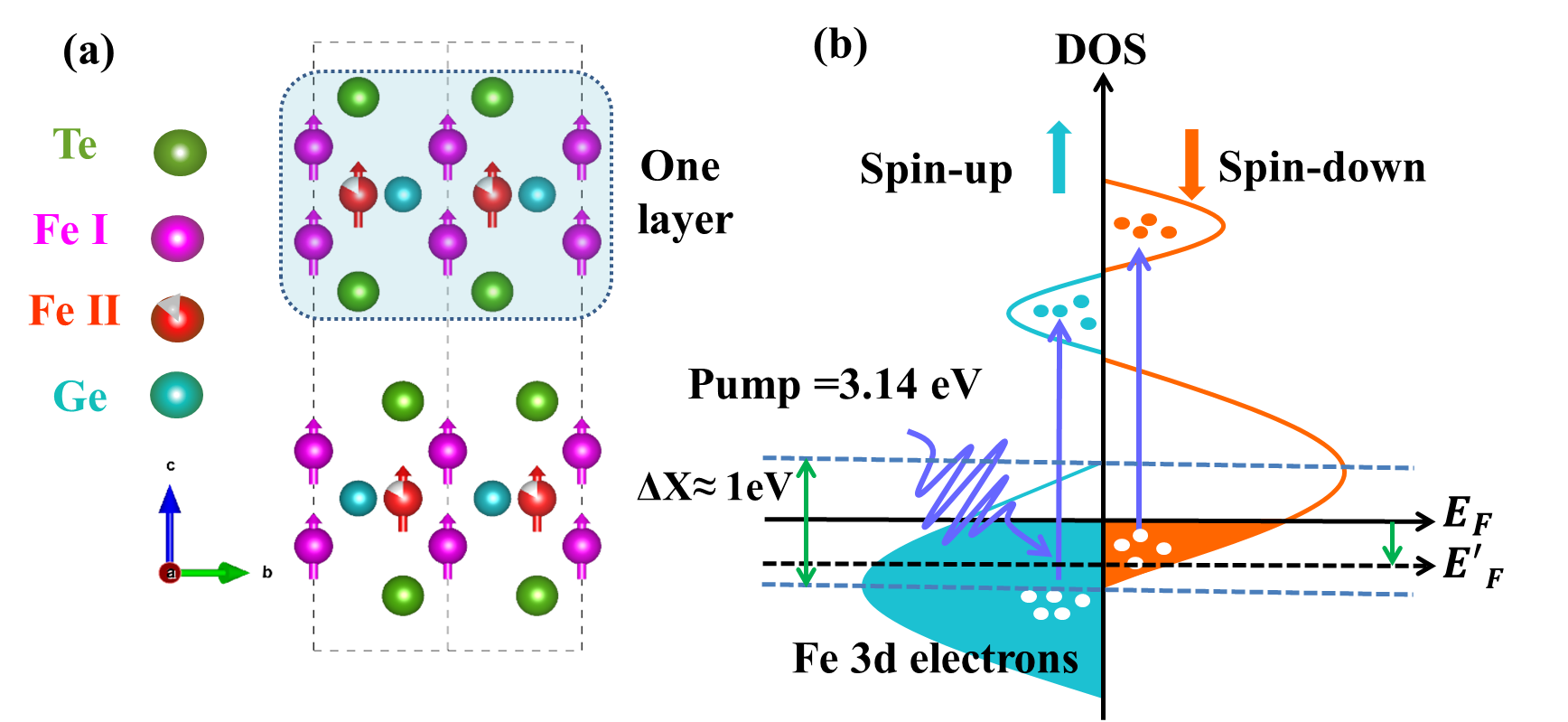}
    \caption{(a) Crystal structure of FGT showing layers stacked along the c-axis. The Fe I (pink) and Fe II (red) atoms occupy two distinct Wyckoff sites, while adjacent Te (green) layers are held together by van der Waals forces. The gray dashed rectangle denotes the unit cell. 
(b) Schematic density of states (DOS) of the Fe $3d$ bands. The spin-up and spin-down bands are exchange split due to spontaneous magnetization. Here, $\Delta_{\mathrm{ex}}$ is the exchange splitting $\approx$ 1 eV \cite{Zhuang2016}. The 3.14 eV pump photon (vertical purple arrows) drives electronic transitions from occupied states below the Fermi level E$_{F}$ to unoccupied states above E$_{F}$.}
    \label{fig_S3_supplementary}
\end{figure}

Fig.~\ref{fig_S3_supplementary} (a) shows the crystal structure of FGT.
The magnetic exchange interactions in FGT arise from a complex interplay between localized and itinerant Fe 3d electrons, characteristic of the correlated metallic ferromagnet \cite{Corasaniti2020}. First-principles analysis has established that the total exchange interactions in FGT arise from the coexistence of strong intralayer and weaker interlayer spin exchange interactions, consistent with its quasi-two-dimensional character \cite{Corasaniti2020, Zhu2023}. However, the ground state of bulk FGT is ferromagnetic. Within each layer, the exchange interactions originate from both direct and super-exchange interactions. Here, direct exchange predominantly occurs between neighboring Fe(I) atoms, while super-exchange interactions act between Fe(I)–Fe(I) and Fe(II)–Fe(II) pairs, mediated by Te and Ge ligands, respectively \cite{Zhu2023}. In addition to these intra- and interlayer interactions, strong uniaxial magnetic anisotropy stabilizes the ferromagnetic order along the crystallographic $c$ axis~\cite{Deng2018}. 

Figure~\ref{fig_S3_supplementary}(b) illustrates a schematic density of states of the Fe $3d$ bands. In the ferromagnetic phase, the spin-up and spin-down bands are exchange split as a consequence of spontaneous magnetization. The corresponding exchange splitting, $\Delta_{\mathrm{ex}}$, is approximately $\sim$1~eV~\cite{Zhuang2016}. Photoexcitation with 3.14~eV pump photons (vertical purple arrows) promotes electrons from occupied states below the Fermi level ($E_F$) to unoccupied states above $E_F$.

\section{Temperature-dependent Decay Amplitudes}

\begin{figure}
    \centering
    \includegraphics[width=1.0\linewidth]{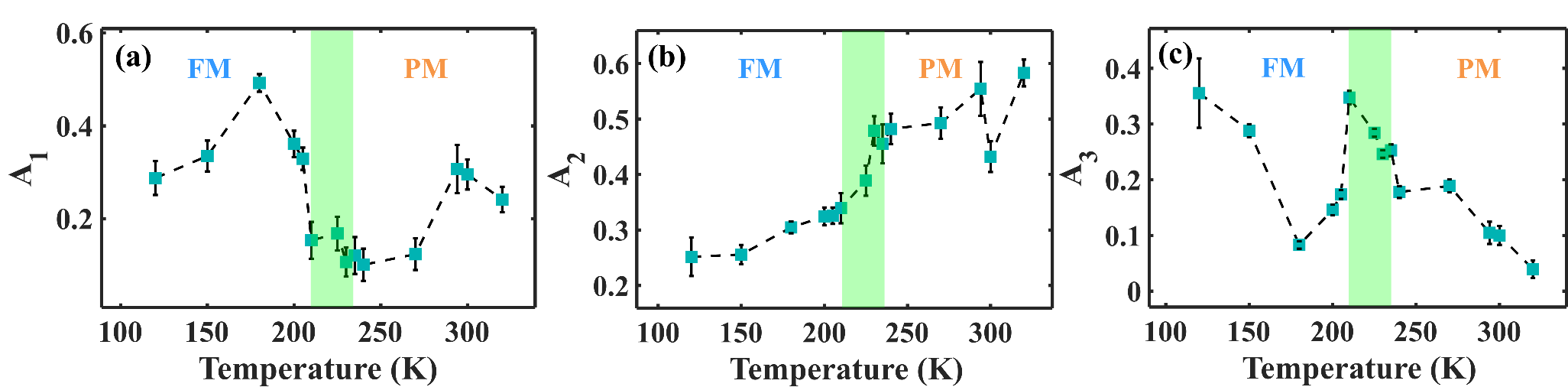}
    \caption{Temperature dependent variation of the decay amplitudes (blue squares) from 120 - 320 K: 			(a) A$_{1}$, (b) A$_{2}$, and (c) A$_{3}$. The green shaded area highlights the region of 						ferromagnetic (FM) to paramagnetic (PM) transition.}
    \label{fig_S4_supplementary}
\end{figure}

Fig. \ref{fig_S4_supplementary} (a)-(c) presents the variation of the electronic decay amplitudes, A$_{1}$, A$_{2}$, and A$_{3}$, as the sample is heated from 120- 320 K. The amplitude A$_{1}$ which represents the strength of the electron-phonon coupling, varies between $\sim$ 10- 50 $\%$ across the measured temperature range. The components A$_{2}$ and A$_{3}$ are attributed to the strength of the interlayer and intralayer spin-lattice coupling, respectively. A$_{2}$ increases from $\sim$ 25 $\%$ to $\sim$ 60 $\%$ with increasing temperature from 120- 320 K. A$_{3}$ varies between  $\sim$ 5 - 40 $\%$ and develops a maximum near the transition temperature.   

\section{Ginzburg-Landau Free energy}

For a second-order phase transition, the Landau free energy can be written as:

\begin{equation}
    f(T)= f_0(T)+ \alpha(T-Tc)p^2+\frac{1}{2}\beta p^4, \label{eq_i}
\end{equation}

where $p$ is the order parameter, and $\alpha$ and $\beta$ are Landau coefficients with $\beta>0$. The term $f_0(T)$ represents the temperature-dependent background of the high-temperature phase in the vicinity of the transition.

The order-parameter dynamics are then described within the relaxational (Model A) framework \cite{Hohenberg1977},

\begin{equation}
    \frac{1}{\mathscr{D}}\frac{\partial p}{\partial t}= -\frac{\partial f}{\partial p} \label{eq_ii}
\end{equation}

where $\mathscr{D}$ denotes the kinetic coefficient governing the relaxation of $p$. Let the equilibrium order parameter be $p_0$, therefore,
\begin{equation}
    \frac{\partial f}{\partial p}\biggr|_{p_0}=0
\end{equation}

For a small fluctuation $\delta p(t)$ in the order parameter, it can be defined as $p(t)=p_0 + \delta p(t)$. Thus, expanding $\frac{\partial f}{\partial p}$ around $p_0$ we get,

\begin{equation}
     \frac{\partial f}{\partial p}= \frac{\partial f}{\partial p}\biggr|_{p_0}+\frac{\partial^2 f}{\partial p^2}\biggr|_{p_0}\delta p+O(\delta p^2) \label{eq_iv}
\end{equation}

Up to linear order in $\delta p$, Eq. \ref{eq_iv} can be rewritten as,
\begin{equation}
    \frac{\partial f}{\partial p}\approx \frac{\partial^2 f}{\partial p^2}\biggr|_{p_0}\delta p
\end{equation}

Substituting in the Eq. \ref{eq_ii}, we get,
\begin{equation}
   \frac{\partial \delta p}{\partial t}=-\mathscr{D}\frac{\partial^2 f}{\partial p^2}\biggr|_{p_0}\delta p
\end{equation}

The solution to the above equation is,

\begin{equation}
    \delta p(t)= \delta p_0 ~e^{-t/\tau} 
\end{equation}
where \begin{equation}
        \frac{1}{\tau}=-\mathscr{D}\frac{\partial^2 f}{\partial p^2}\biggr|_{p_0}\label{eq_viii}
\end{equation}

Hence, $\tau$ can be estimated from the second-order derivative of free energy, which can be obtained from Eq. \ref{eq_i} as follows:

\begin{equation}
    \frac{\partial^2 f}{\partial p^2}= 2\alpha (T-T_c)+6\beta p^2
\end{equation}

At equilibrium, $\frac{\partial f}{\partial p}\biggr|_{p_0}=0$, which gives $p_0=0$ for $T>T_c$ and $p_0^2= \frac{-2\alpha (T-T_c)}{2\beta}$ for $T<T_c$. Thus, we have

\begin{equation}
    \frac{\partial^2 f}{\partial p^2}\biggr|_{p_0}= 2\alpha (T-T_c) ~~~ \text{for}~~ T>T_c
\end{equation}

And substituting it into Eq. \ref{eq_viii} gives,

\begin{equation}
    \tau=\frac{1}{2\alpha \mathscr{D}(T-T_c)}
\end{equation}

\begin{equation}
    \tau=\tau_{0}^+(T-T_c)^{-1} ~~\text{where}~~ \tau_{0}^+= (2\alpha\mathscr{D})^{-1}
\end{equation}

For $T<T_c$,
\begin{equation}
    \frac{\partial^2 f}{\partial p^2}\biggr|_{p_0}= 2\alpha (T-T_c)+6\beta p_0^2
\end{equation}

\begin{equation}
\tau=\tau_{0}^-(T_c-T)^{-1} ~~\text{where}~~ \tau_{0}^-= (4\alpha\mathscr{D})^{-1}
\end{equation}

Therefore,
\begin{equation}
\tau=\begin{cases} \frac{1}{2\alpha\mathscr{D}(T-T_c)} & \text{for } T>T_c \\
\frac{1}{4\alpha\mathscr{D}(T_c-T)} & \text{for } T<T_c\end{cases}
\end{equation}
 
Thus, $\tau$ exhibits critical slowing down in the vicinity of T$_c$, which is experimentally manifested in the slowest relaxation component, $\tau_3$. The corresponding critical amplitudes above and below T$_c$, denoted as $\tau_0^{+}$ and $\tau_0^{-}$, respectively, define the amplitude ratio $R_{\tau} = \tau_0^{+}/\tau_0^{-}$. While conventional scaling predicts $R_{\tau} \simeq 2$, we instead obtain $R_{\tau} \simeq 0.6$ (see main text). This anomalous amplitude ratio indicates that the relaxation dynamics associated with $\tau_3$ deviates from simple universal scaling behavior.

\section{Data fitting}

The experimental data were fitted using the Nelder-Mead simplex algorithm, implemented in MATLAB via the fminsearch function. The time- resolved reflectivity data was fitted using a tri-exponential function multiplied by an error function to account for the finite rise time, as defined in Eq.(1) of the main manuscript. The fit was obtained by minimizing the residual S= $\Delta$R/R- $\Delta$R/R$_{\text{tri-exp}}$. Error bars were estimated using the orthogonal distance regression (ODR) method \cite{Boggs1992}. In  MATLAB, this procedure involved computing the Jacobian matrix and subsequently the variance-covariance matrix to extract standard errors of the fitted parameters.

\begin{figure}
   \centering
  	\includegraphics[width=1\linewidth]{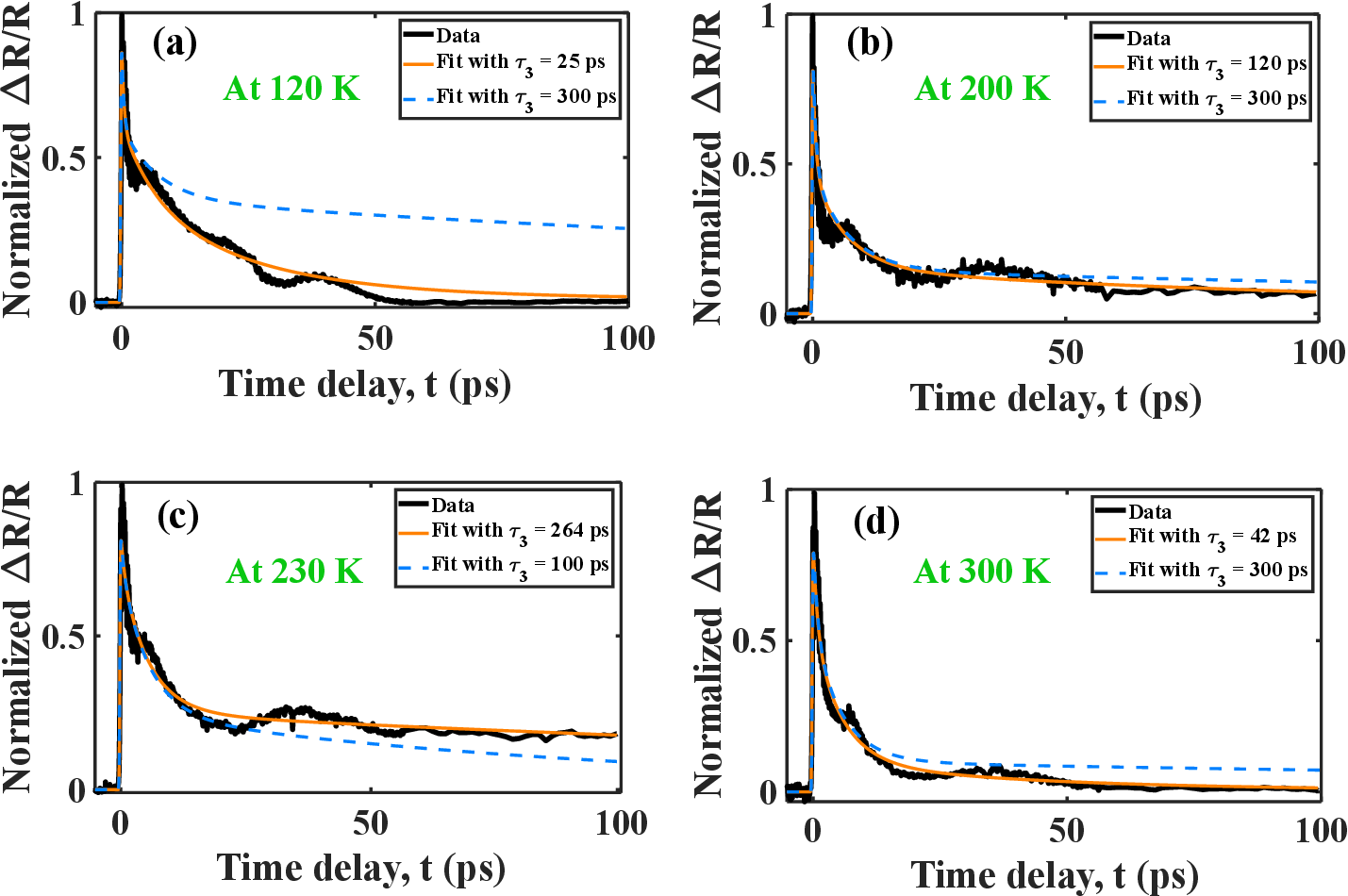}
  	\caption{The time-resolved reflectivity with the fit at four different temperatures, (a) 120 K, (b) 200 K, (c) 230 K, and (d) 300 K. Different values of $\tau_{3}$ were used to check the best fit to the data.}
    \label{fig_S5_supplementary}
\end{figure}

To assess the robustness of the fitting procedure, the time-resolved reflectivity traces at four different temperatures, 120 K, 200 K, 230 K, and 300 K were refitted using a range of imposed values of the slowest decay constant, $\tau_{3}$, keeping the other parameters fixed (Fig. \ref{fig_S5_supplementary}). We find that only near the magnetic transition (200 K and 230 K) does the fit require a comparatively larger $\tau_{3}$ [Figs. \ref{fig_S5_supplementary}(b) and (c)], whereas at 120 K and 300 K the data are best reproduced with smaller $\tau_{3}$ values [Figs. \ref{fig_S5_supplementary}(a) and (d)]. Fits using $\tau_{3}$ values either smaller or larger than the optimal choice show clear deviations from the experimental traces. The final $\tau_{3}$ values reported in the main text correspond to those that minimize these deviations.

\section{Anharmonic decay}

Above T$_{c}$, the phonon mode frequency ($\nu$) and the linewidth ($\Gamma$) follow the cubic anharmonic decay in agreement with earlier Raman measurements \cite{Du2017}, and are given by the following equations \cite{Cardona1983,Klemens1966}:

\begin{subequations}
\begin{equation}
 \nu(T) = \nu_{0}-A\left[1+\frac{2}{\text{exp}\left(\frac{h\nu_{0}}{ 2k_{B}T}\right)-1}\right]  
 \label{eq3a}
\end{equation}

\begin{equation}
 \Gamma(T) = \Gamma_{0} + C\left[1+\frac{2}{\text{exp}\left(\frac{h\nu_{0}}{ 2k_{B}T}\right)-1}\right]  \label{eq3b}
\end{equation}
\end{subequations}

Here, $ h$ is the Planck’s constant and $k_{B}$ is the Boltzmann constant. In Eq. (\ref{eq3a}), $\nu_{0}$ denotes the phonon frequency at zero temperature, while the second term represents the intrinsic anharmonic contribution arising from the decay of an optical phonon into two acoustic phonons of equal energy and opposite momenta, corresponding to the real part of the phonon self-energy \cite{Cardona1983,Klemens1966}. In Eq. (\ref{eq3b}),  $\Gamma_{0}$ is the disorder-induced temperature-independent linewidth of the phonon at absolute zero temperature, whereas the second term is anharmonic contribution originating from the imaginary part of the phonon self-energy \cite{Cardona1983,Klemens1966}. The solid red line in Fig. 4(c) and 4(d) in the main manuscript is the fit using the anharmonic decay model and the parameters obtained after fitting are: $ \nu_{0} $ = 3.78 $ \pm $ 0.02 THz, $ A $ = -0.024  $ \pm $ 0.004 THz, $  \Gamma_{0} $ = 0.04 $ \pm $ 0.03 THz  and $ C $ = 0.036 $ \pm $ 0.005 THz.

\section{Fano parameter with fluence}

\begin{figure*}
    \centering
    \includegraphics[width=0.9\linewidth]{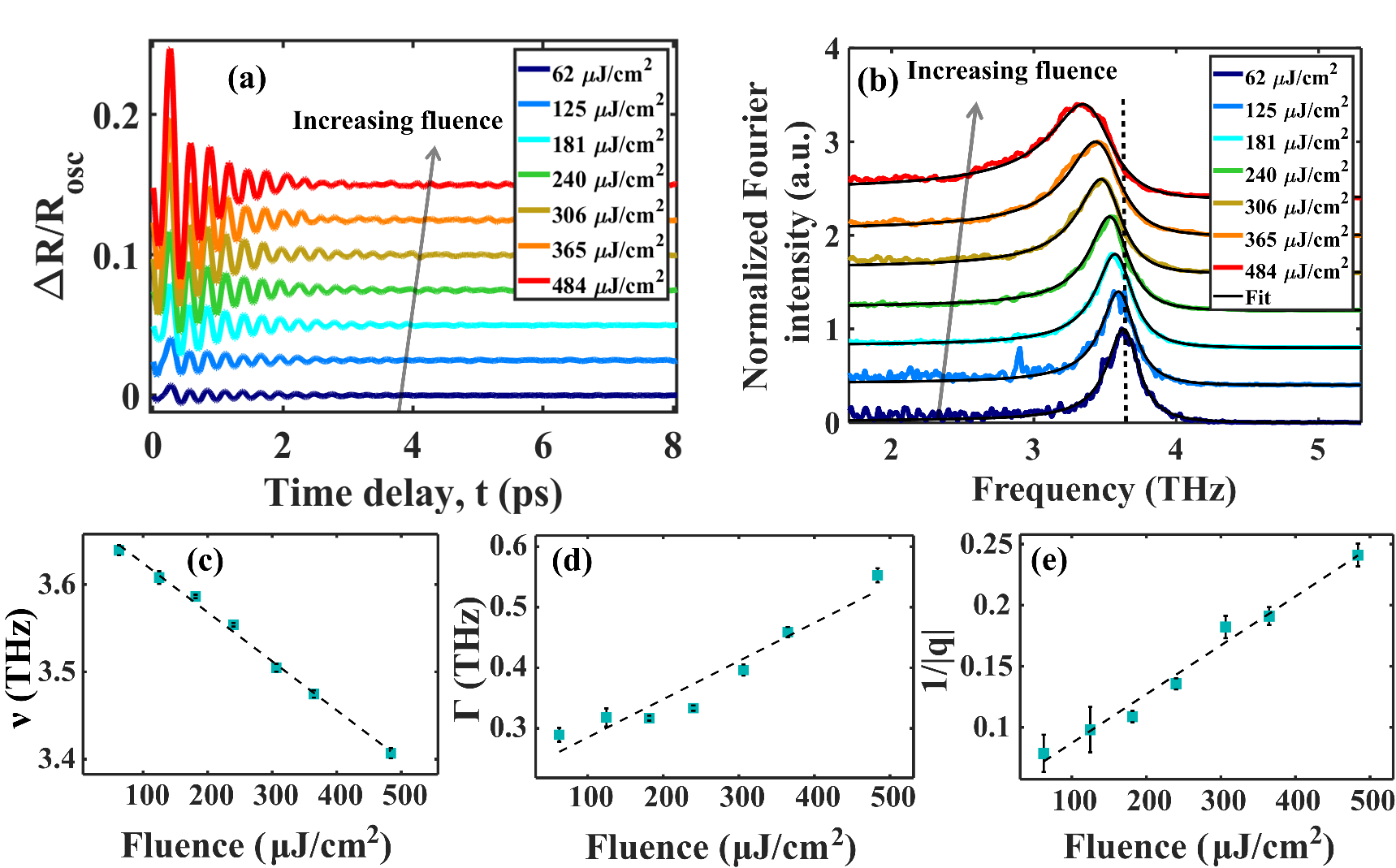}
    \caption{ (a) The oscillatory component extracted from the transient $\Delta$R/R signal and the corresponding (b) Normalized Fourier intensity spectra as a function of fluence. The black solid line in (b) represents a fit to the Fano function (Eq.2 of main manuscript). The arrow indicated the direction of increasing fluence and the black dotted lines serve as guide to the eye. The phonon frequency, linewidth, and Fano asymmetry parameter extracted from these fits are summarized in (c)-(e) respectively.  The black dashed line in the (c)-(d) correspond to linear fits.  
}    \label{fig_S6_supplementary}
\end{figure*}

Figs. \ref{fig_S6_supplementary} (a) and (b) present the fluence evolution of the optical phonon and the normalized Fourier intensity, respectively.  The Fourier intensity is fitted with the Fano function (Eq. 2 of the main manuscript) and the extracted parameters are displayed in Fig. \ref{fig_S6_supplementary}(c)-(e). 
The absolute value of the Fano parameter increases nearly linearly with pump fluence, indicating that the coupling between the electronic continuum and the phonon is strongly enhanced at higher excitation carrier densities  (Fig. \ref{fig_S6_supplementary}(e)). A plausible explanation is that increasing fluence enhances the carrier density near the Fermi level, thereby extending the electronic continuum available for interference.  
The phonon frequency exhibits a clear softening (see Fig. \ref{fig_S6_supplementary}(c)) due to the dense population of photoexcited carriers that transiently screen interatomic forces and weaken the effective force constants. In addition, since higher fluence generates a larger non-equilibrium phonon population, it opens additional decay channels for the phonons that shorten the phonon lifetime and lead to linewidth broadening (see Fig. \ref{fig_S6_supplementary}(d)).

\section{Acoustic strain pulse dynamics}

\begin{figure*}
    \centering
    \includegraphics[width=0.8\linewidth]{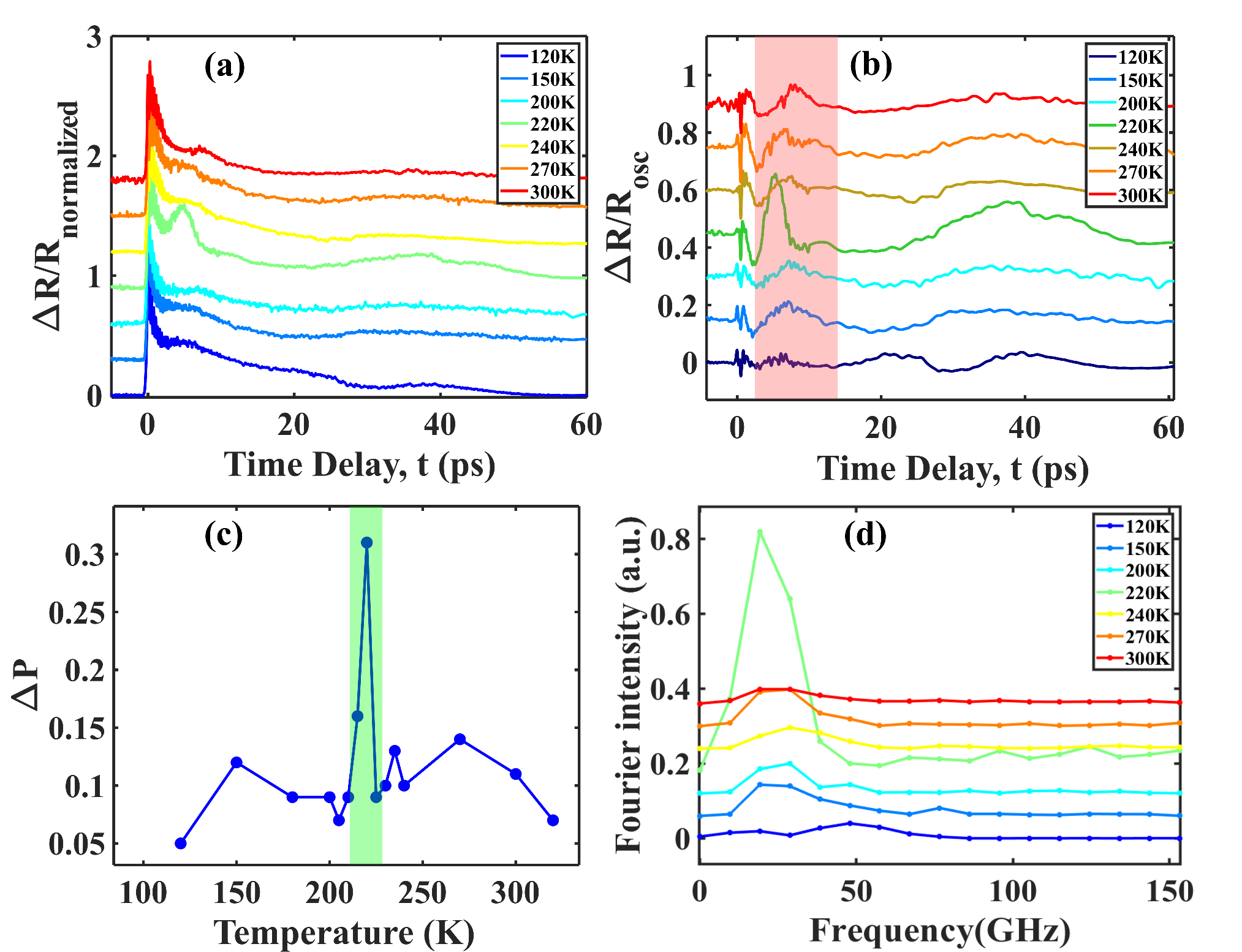}
    \caption{ The transient normalized $\Delta$R/R data at few selected temperatures from 120-320 K. (b) The strain component extracted from the transient $\Delta$R/R signal is shown at few selected temperatures from 120-320 K. The red shaded region highlights the strain whose amplitude (dip to peak difference ($\Delta$P)) is plotted in the (c) as a function of temperature. The Fourier intensity spectra with increasing temperature are presented in (d), respectively. 
    }    \label{fig_S7_supplementary}
\end{figure*}

Next, we focus on the temperature evolution of the acoustic strain pulse. Fig. \ref{fig_S7_supplementary} displays the normalized data ( Fig. \ref{fig_S7_supplementary} (a)) and extracted strain pulse (Fig. \ref{fig_S7_supplementary} (b)) obtained after removing the triexponential background and filtering out the optical phonon component (using a Butterworth filter in MATLAB) and the resultant strain curves were smoothed using Locally Weighted Scatterplot Smoothing (LOWESS) implemented in MATLAB. Low-frequency oscillations corresponding to a accoustic strain pulse are clearly resolved, with an amplitude that increases and becomes chirped as the pulse propagates. When a femtosecond laser pulse irradiates a magnetic material, it can induce strain pulses through several mechanisms, including thermoelastic stress, deformation potential, and magnetostriction \cite{Thomsen1986, Reppert2020, Chauhan2025}. The rapid energy deposition by the pump pulse causes a sudden lattice expansion and generates a propagating bipolar strain pulse, typically consisting of a leading compressive component followed by a trailing tensile part \cite{Anjan2025}. The amplitude and shape of these strain pulses are strongly influenced by the material’s elastic constants, optical absorption, and magnetic ordering, providing a sensitive probe of lattice-spin interactions on ultrafast timescales. 

The strain pulse, launched by the pump pulse and detected by the probe reflected from both the sample surface and the propagating strain front, exhibits a distinctive temperature dependence: its amplitude remains nearly constant across most temperatures but increases sharply near the ferromagnetic transition. The red-shaded region in Fig. \ref{fig_S7_supplementary} (b) marks the portion of the signal used to extract the dip-peak difference ($\Delta$P), shown in Fig. \ref{fig_S7_supplementary}(c), which peaks near the transition temperature. A similar enhancement is seen in the Fourier intensity spectra [Fig. \ref{fig_S7_supplementary}(d)], which display a broad maximum at $\sim$ 28 GHz. These features indicate strong coupling between the strain pulse and magnetic order, evidencing pronounced magnetoelastic interactions that link the magnetic subsystem to the transiently strained lattice. Similar behaviour of the acoustic strain has also been reported previously in the ferrimagnetic insulator GdTiO$_{3}$ \cite{Lovinger2020}. From the observed acoustic strain frequency, one can estimate the sound velocity ($v_{s}$) in FGT from the relation $v_{s} = \lambda f / (2n)$, where $\lambda$ is the probe wavelength, $f$ is the oscillation frequency, and $n$ is the refractive index. Using $\lambda \approx 790$ nm, $f \approx 28$ GHz, and $n \approx 4$ for a similar vander waal magnetic material Cr$_{2}$Ge$_{2}$Te$_{6}$ \cite{Idzuchi2023}, we obtain $v_{s} \approx 2.8$ $\times$ 10$^{3}$ m/s, similar to the value observed in layered magnetic materials like CrSiTe$_{3}$ \cite{Anjan2025} and MPS$_{3}$ compounds \cite{Wright2024}.

\begin{figure}
   \centering
  	\includegraphics[width=1\linewidth]{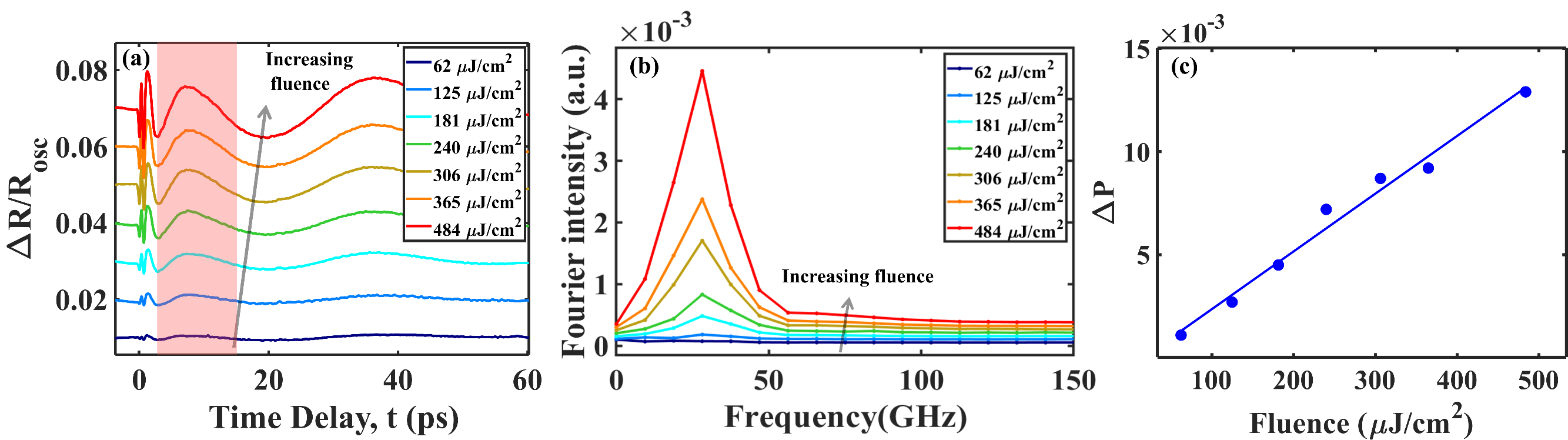}
  	\caption{Fluence-dependent variation of the strain pulse: (a) The extracted strain pulse after removing 	the exponential background. And the corresponding (b) Fourier intensity with increasing fluence. (c) The strain pulse amplitude as a function of fluence ($\Delta$P = dip-to-peak difference of the shaded area in (a)). }
    \label{fig_S8_supplementary}
\end{figure}

Figs. \ref{fig_S8_supplementary} (a) and (b) display the extracted strain pulse and the corresponding Fourier intensity as a function of fluence. The amplitude of the strain pulse (Fig. \ref{fig_S8_supplementary} (c) dip to peak of $\Delta$R/R$_{osc}$ shaded region) and Fourier intensity increases linearly with the increase in pump fluence. However, the temporal profile of the strain pulse and bandwidth of the normalized FFT intensity of the strain pulse remains largely unaffected with increase in pump fluence, indicating that the strain pulse dynamics are robust against variations in excitation density.

\section{Thermal Phonon-Enhanced Fano Asymmetry in the Paramagnetic Phase of $\mathrm{\textbf{Fe}}_3\mathrm{\textbf{GeTe}}_2$}

\subsection{Microscopic Hamiltonian and Interaction Pathways}
The emergence and subsequent thermal boosting of the Fano asymmetry are modeled via a Hamiltonian that explicitly treats the interaction between the zone-center optical mode, the itinerant electronic subsystem, and the acoustic phonon bath. We define $\mathbf{q}_{probe}$ as the momentum transfer provided by the optical probe, which in our geometry is restricted to the long-wavelength limit ($|\mathbf{q}_{probe}| \approx 10^{-3} \text{\AA}^{-1}$). The total Hamiltonian $H$ is:
\begin{equation}
H = H_e + H_{\text{opt}} + H_{\text{ac}} + H_{e\text{-opt}} + H_{\text{anh}} + H_{e\text{-ac}}
\end{equation}
where $H_e = \sum_{\mathbf{k}\sigma} \epsilon_{\mathbf{k}} c_{\mathbf{k}\sigma}^\dagger c_{\mathbf{k}\sigma}$ describes the itinerant $d$-band dispersion ($\epsilon_\mathbf{k} = \alpha|\mathbf{k}|^2-\mu$). $H_{\text{opt}} = \nu_0 \sum_{\mathbf{q}} b_{\mathbf{q}}^\dagger b_{\mathbf{q}} $ and $H_{\text{ac}} = \sum_{\mathbf{p}, \nu} \nu_{\mathbf{p}, \nu} a_{\mathbf{p}, \nu}^\dagger a_{\mathbf{p}, \nu} $ are the free optical and acoustic phonon Hamiltonians, respectively. Here, $\nu_0$ is the momentum-independent optical frequency and $\nu_{\mathbf{p}, \nu}$ are the momentum-dependent acoustic frequencies. The interaction terms are:
\begin{align}
H_{e\text{-opt}} &= \frac{g}{\sqrt{N}} \sum_{\mathbf{k},\mathbf{q},\sigma} c_{\mathbf{k}+\mathbf{q},\sigma}^\dagger c_{\mathbf{k}\sigma} (b_{\mathbf{q}} + b_{-\mathbf{q}}^\dagger) \\
H_{\text{anh}} &= \sum_{\mathbf{p}, \nu} \mathcal{V}_{\text{anh}} (b_{\mathbf{q}} a_{\mathbf{p}, \nu}^\dagger a_{\mathbf{q}-\mathbf{p}, \nu}^\dagger + \text{H.c.}) \\
H_{e\text{-ac}} &= \sum_{\mathbf{k}, \mathbf{p}, \sigma, \nu} g_{\text{ac}}(\mathbf{p}, \nu) c_{\mathbf{k}+\mathbf{p}, \sigma}^\dagger c_{\mathbf{k}, \sigma} \left( a_{\mathbf{p}, \nu} + a_{-\mathbf{p}, \nu}^\dagger \right)
\end{align}
where $g$ is the bare electron-optical phonon coupling, $\mathcal{V}_{\text{anh}}$ is the third-order anharmonic coupling (Klemens decay), and $g_{\text{ac}}$ is the electron-acoustic phonon coupling.

We study this system in a two-temperature setup: the electron continuum (the incoherent electron-hole states) is at a transient temperature $T_e$, while the phonons remain at a lower temperature $T_p$, close to the base sample temperature. $H_{e\text{-opt}}$ provides the baseline Fano asymmetry due to the thermal smearing of the Fermi surface at $T_e$, which ensures the on-shell condition required for the relaxation of excited electrons. 

However, this direct channel is kinematically suppressed because the optical probe only accesses momenta near the zone center ($|\mathbf{q}_{probe}| \to 0$). Furthermore, a purely electronic Fano effect would typically decrease with increasing $T_p$ due to dephasing, which contradicts our experimental observations. Instead, we find that the anharmonic Klemens decay ($H_{\text{anh}}$) where the optical phonon at $\mathbf{q}$ decays into acoustic phonons with momenta ($\mathbf{p}$ and $\mathbf{q}-\mathbf{p} $), is the key mechanism. This pathway generates an increasing asymmetry with $T_p$ by facilitating a broad spectral overlap with the pump-induced electronic continuum.

\subsection{Effective Green's Function and Optical Intensity}

To obtain the effective dynamics of the optical mode, we formally integrate out the acoustic phonon degrees of freedom. This procedure generates a complex self-energy $\Sigma_{\text{anh}}(\nu, T_p)$ for the optical phonon. The real and imaginary parts correspond to the temperature-dependent frequency shift $\Delta \nu(T_p)$ and the anharmonic broadening $\Gamma_{\text{anh}}(T_p)$:
\begin{align}
\nu(T_p) &= \nu_0 + \text{Re}[\Sigma_{\text{anh}}] \\
\Gamma_{\text{anh}}(T_p) &= \text{Im}[\Sigma_{\text{anh}}]
\end{align}
Simultaneously, the integration of $H_{\text{ac}}$ in the presence of $H_{e\text{-ac}}$ generates a renormalized coupling $V(T_p) = g + \delta V_{\text{ac}}(T_p)$. While the vertex correction increases proportional to $T_P$, given that vertex corrections are typically suppressed by Migdal's theorem in the adiabatic limit, we neglect $\delta V_{\text{ac}}$ and focus on the self-energy effects.

The coupled system is represented by the inverse Green's function matrix $\mathbf{G}^{-1}$ in the basis of phonon coordinates and electronic density fluctuations, $\psi_\mathbf{q} = (x_\mathbf{q}, \rho_\mathbf{q})^T$:
\begin{equation}
\mathbf{G}^{-1}(\mathbf{q}, \nu; T_p, T_e) = \begin{pmatrix} [D^R(\mathbf{q}, \nu, T_p)]^{-1} & g \\ g & -[\Pi_0^R(\mathbf{q}, \nu, T_e)]^{-1} \end{pmatrix}
\end{equation}
where $D^R$ is the phonon propagator dressed by $\Sigma_{\text{anh}}$, and $\Pi_0^R$ is the electronic polarization bubble at $T_e$. The resulting optical intensity $I(\nu)$ captures the quantum interference:
\begin{equation}
I(\nu) \propto \operatorname{Im}\Bigg[ \mathcal{M}_e^2 \Pi_0^R + \frac{D^R \left( \mathcal{M}_p + g \mathcal{M}_e \Pi_0^R \right)^2}{1 - g^2 D^R \Pi_0^R} \Bigg]
\end{equation}
where $\mathcal{M}_{e}$ and $\mathcal{M}_{p}$ are the respective bare electronic and phonon dipole matrix elements.

\subsection{Thermal Phonon-Enhanced Fano Asymmetry}
For the numerical study, we assume a parabolic electronic dispersion with $\alpha=410$~meV \cite{Kim2018} and $\mu$ fixed by the experimental Fermi \cite{Rana2022} ($3.8\times10^5$~m/s). We vary the effective coupling $g/\alpha$ between $0.1$ and $0.6$, which is within estimates found in literature \cite{Badrtdinov2023}. The real and imaginary parts of  $\Sigma_{anh}$ are identified with Eq. 17 (a) and (b).  The Klemens parameters $A$ and $C$ ( see, Eq. (\ref{eq3a}) and (\ref{eq3b})) are calibrated to match the observed softening and broadening at $g=0$. 

We find a finite baseline Fano asymmetry even for zero Klemens coupling ($A=C=0$), confirming that the high $T_e$ provides sufficient spectral overlap for relaxation. However, switching on the Klemens parameters causes the softening and broadening to grow monotonically with $T_p$. The resulting Bose-enhanced population of acoustic phonons facilitates momentum transfer $|\mathbf{p}|$ across the entire Brillouin zone. 

According to standard theory, the on-shell condition for a parabolic electronic band requires that $2\alpha |\mathbf{k}||\mathbf{q}|=\nu_0$, where the electron-phonon interaction scatters an electron from state $\mathbf{k}$ to $\mathbf{k}+\mathbf{q}$ by absorbing a phonon with momentum $\mathbf{q}$ and frequency $\nu_0$. Since the optical probe is limited to accessing momenta close to the zone center ($|\mathbf{q}_{\text{probe}}| \approx 10^{-3} \text{ \AA}^{-1}$), this kinematic constraint can only be satisfied by electronic states with a sufficiently large momentum $|\mathbf{k}|$. At low electron temperatures, within a simplified one-band model, the maximum available momentum is bounded by the Fermi momentum ($|\mathbf{k}_{\text{max}}| \approx |\mathbf{k}_F|$), which is insufficient to satisfy the kinematic relation. Consequently, a large transient electronic temperature ($T_e$) is crucial, as it enables the thermal occupation of states far above $k_F$, thereby opening the necessary scattering channels. We note that this remains a qualitative argument; in a more realistic multi-band setting, inter-band transitions may further help bridge this energy gap and could potentially activate the relaxation process even at lower temperatures.
In either case, while the small probe momentum $|\mathbf{q}_{\text{probe}}|$ imposes strict constraints on the accessible electronic momenta, the intermediate \textit{virtual acoustic phonons bridge the kinematic gap efficiently}. As these thermally populated acoustic modes have access to momenta spanning the entire Brillouin zone, they facilitate a much broader spectral overlap between the discrete optical mode and the electronic continuum. In this way, the acoustic-mediated pathway systematically enhances the interaction, thereby boosting the Fano asymmetry as a function of the lattice temperature ($T_p$).

Finally, below $T_c$, the onset of ferromagnetic order induces a large exchange splitting. As the $A_{1g}$ optical phonon cannot mediate the spin-flip/angular momentum transfer required for scattering between these split bands, the scattering channel is quenched, leading to the observed suppression of Fano asymmetry in the ordered phase, regardless of the kinematic constraints as discussed above.

\end{document}